\documentclass[prd,eqsecnum,nofootinbib,preprint,floatfix,finalold,superscriptaddress]{revtex4-1}

\usepackage{colordvi}
\usepackage{graphicx}
\usepackage{subfigure}
\usepackage{amsmath}
\usepackage{amsfonts}

\reversemarginpar
\def\<{\langle}
\def\>{\rangle}
\def\Leff{L_{\rm eff}}
\def\Neff{N_{\rm eff}}
\def\Tr{{\rm Tr}\,}
\def\tr{{\rm tr}\,}
\def\Eq#1{Eq.~(\ref{#1})}

\renewcommand\Re{{\rm Re}\,}

\def\Ref#1{Ref.~\cite{#1}}

\begin{document}

\pagenumbering{roman}

\title{Large-$N$ reduction in QCD with two adjoint Dirac fermions}

\author{Barak Bringoltz}
\affiliation{
\mbox{IIAR -- the Israeli Institute for Advanced Research, Rehovot, Israel}\\}
\affiliation{Department of Physics,
University of Washington, Seattle, WA 98195-1560, USA}
\author{Mateusz Koren}
\affiliation{M. Smoluchowski Institute of Physics,\\
\mbox{Jagiellonian University, Reymonta 4, 30-059 Cracow, Poland}\\}
\affiliation{Department of Physics,
University of Washington, Seattle, WA 98195-1560, USA}
\author{Stephen R.~Sharpe}
\affiliation{Department of Physics,
University of Washington, Seattle, WA 98195-1560, USA}

\date{\today}

\begin{abstract}
We use lattice simulations to study the single-site version of 
$SU(N)$ lattice gauge theory with two flavors of Wilson-Dirac
fermions in the adjoint representation,
a theory whose large volume correspondent is
expected to be conformal or nearly conformal.  Working with $N$ as
large as $53$, we map out the phase diagram in the plane of
bare `t Hooft coupling, $g^2N$, and of the lattice quark mass, $am$, 
and look for the region where the $Z_N^4$
center symmetry of the theory is intact. In this
region one expects the large-$N$ equivalence of the single site
and infinite volume theories to be valid. As for the
$N_f=1$ case (see Phys.~Rev.~D~{\bf 80}:~065031), 
we find that the center-symmetric region is
large and includes both light fermion masses and masses at
the cutoff scale. We study the $N$-dependence of the width of this
region and find strong evidence that it remains
of finite width as $N\to\infty$.
Simulating with couplings as small as 
$g^2N = 0.005$, we find that the width shrinks slowly with decreasing
$g^2N$, at a rate consistent with analytic arguments.
Within the center-symmetric region
our results for the phase structure, when extrapolated to $N=\infty$,
 apply also for the large volume theory, which is
minimal walking technicolor at $N=\infty$.
We find a first-order transition as a function of $am$
for all values of $b$, 
which we argue favors that the theory is confining in the infrared.
Finally, we measure the eigenvalue densities of the Wilson-Dirac operator
and its hermitian version, and use large Wilson loops to study the
utility of reduction for extracting physical observables.

\end{abstract}

\maketitle

\setcounter{page}{1}
\pagenumbering{arabic}
\pagestyle{plain}

\section{Introduction}
\label{intro}
There has been a recent revival of interest in the possibility of
using complete volume reduction for the infinite $N$ (number of
colors) limit of QCD and QCD-like theories. If this reduction
holds, then the theory, defined nonperturbatively on a lattice,
gives predictions that, at infinite $N$,
are independent of the number of sites. Specifically
this means that the theory defined on a single site, or a small, fixed
number of sites, is large-$N$ equivalent to the corresponding infinite
volume theory with the same bare parameters~\cite{EK}.

Reduction to a single-site has long been known to fail for the
pure gauge theory (and thus also for QCD in 't Hooft's large-$N$
limit, since quark contributions are suppressed by $1/N$
in this limit)~\cite{BHN,KM,Okawa}.
This failure is due to the breakdown
of one of the conditions needed for a large-$N$ orbifold equivalence
between the single-site and large volume theories 
(see Refs.~\cite{Neuberger2002} and \cite{KUY1}).
This key condition is that the $Z_N^4$ center symmetry of the
single-site theory must be unbroken.
This symmetry breaks spontaneously\footnote{
Strictly speaking, the symmetry is spontaneously broken only for
$N\to\infty$. In practice, however, effective spontaneous symmetry
breaking is seen in simulations at finite but large values of $N$, and we use
the terminology of phase transitions throughout this article.}
in the single-site pure gauge theory (the Eguchi-Kawai [EK] model~\cite{EK}).
This is expected from perturbation theory (PT), where, at leading order,
the effective potential for eigenvalues of the holonomy around the 
compact direction (the Polyakov loop) leads to attraction 
and thus clumping~\cite{BHN,KM}.
Several years ago, it was realized that the addition of massless fermions 
that reside in the
adjoint representation and that have periodic boundary conditions 
in the compact directions leads (in perturbation theory)
to a repulsion between eigenvalues,
which in turn leads to a uniform distribution 
of these eigenvalues~\cite{KUY2}. 
In this case the center symmetry is unbroken and reduction holds.

Two of us have previously investigated the single-site theory with 
a single Dirac adjoint fermion (discretized using Wilson fermions),
finding that, for small to rather large values of the inverse
't Hooft coupling, $b=1/(g^2 N)\in[0,1]$, there is a large range of
values of the quark mass for which the center symmetry 
appears to be unbroken~\cite{AEK}.  This result was unanticipated
because several leading-order perturbative calculations, done with a
single compact direction (or a single site along only one Euclidean
direction) show that the symmetry breaks if the physical mass $m$
exceeds a value of $O(1/aN)$~\cite{BBPT,HollowoodMyers2009,AHUY}.
As $m$ increases from $\sim 1/aN$ to $\sim 1/a$, 
the perturbative calculations indicate that the eigenvalue density 
of the link in the short direction will  
form a number of clumps, starting with
$O(N)$ clumps at very small masses, 
and decreasing to a single clump at infinite mass.
The results of Ref.~\cite{AEK} (which used $N$ up to 15) 
do not follow this pattern,
instead finding no clumping
for masses up to of $O(1/a)$ for all values of $N$.
A semi-quantitative understanding of these results has recently
been given in Refs.~\cite{AHUY,UY10}.
With more than two compact dimensions, the 
fluctuations in the eigenvalues can overwhelm the tendency to clump,
and this happens up to masses of $O(1/a)$.
The numerical results for the phase diagram obtained in Ref.~\cite{AEK}
have also been checked, and extended, in Ref.~\cite{AHUY}. 
We also note that simulations with
$N_f=1/2$, $1$ and $2$ massless overlap adjoint fermions
also find no center-symmetry breaking, at least at 
large $b$~\cite{HN1,HN2}.

In the present paper we extend our investigations to $N_f=2$.
The main motivation for doing so is that the corresponding
infinite volume theory is expected to be nearly conformal,
and thus a candidate ``walking technicolor'' model.\footnote{%
For recent reviews of technicolor models on the lattice,
see Refs.~\cite{Rummukainen:2011xv} and \cite{DelDebbio:2011rc}.
}
Indeed the theory with two colors  is the theory with the smallest
field content that lies close to the conformal window,
and has been dubbed the ``minimal walking technicolor'' (MWT) model.
If reduction holds, then we should be able to study a close
relative of MWT, i.e. the theory with $N=\infty$.
One naively expects only a weak dependence on $N$, 
because the number of both gluonic and
fermionic degrees of freedom scale as $N^2$.

We also note that the $N_f=2$ AEK (Adjoint Eguchi-Kawai) model
is related by a combination of orbifold and orientifold equivalences
to the QCD-like theory with $2N_f=4$ fermions 
in the two index anti-symmetric (AS) irrep~\cite{ASV,KUY2}.\footnote{%
This equivalence holds only in charge-conjugation even subsectors.}
This theory in turn is the large-$N$ limit of QCD with $2N_f$ quarks,
but with the limit taken with the quarks in the AS irrep
(which is equivalent to the anti-fundamental for $N=3$).
This is the Corrigan-Ramond large-$N$ limit of QCD~\cite{CorriganRamond},
which differs from the 't Hooft limit in having fermion loops.\footnote{%
Note that here
taking the Corrigan-Ramond limit moves one from a theory
which is not close to the conformal window 
(4 quarks in the fundamental irrep)
to one that is (4 quarks in the AS irrep),
suggesting that $1/N$ corrections are probably large, 
at least in this respect.}

Our main effort herein is to determine the phase structure of
the $N_f=2$ AEK model. 
To do so we have upgraded our simulation algorithm from a
Metropolis algorithm, with CPU scaling as $N^{6-8}$, to a
Hybrid Monte-Carlo (HMC) algorithm, for which we find
${\rm CPU}\propto N^{4-4.5}$.
This allows us to reach much larger values of $N$, and
to improve the statistics.
Together, these advances allow us to study
the nature of the symmetry breaking
in more detail than in the $N_f=1$ study~\cite{AEK},
allowing us to compare with the theoretical expectations of
Refs.~\cite{AHUY,UY10}.
Our main result is that we
find the phase diagram to be qualitatively similar to
that for $N_f=1$, with center symmetry remaining
unbroken for masses up to $O(1/a)$.
Specifically, our evidence suggests that, at fixed coupling,
although the range of masses for which
the symmetry is unbroken shrinks somewhat as $N$ increases,
it remains of width $\sim 1/a$ as $N\to\infty$.
Our strongest evidence for this is at $b=1$, but our results
suggest that this holds for $b=0-200$, i.e. for the entire range
of coupling that one could possibly be interested in.
Thus our results suggest that
one can use adjoint fermions
of almost any mass to ``stabilize'' reduction. This is only
expected to fail in the extreme weak coupling limit ($b\to\infty$).

This is an encouraging result, and so we have made the
first steps in trying to see if reduction can be used
to obtain results for physical quantities. 
The key question is how large a value of $N$ is needed
so that the physical contributions are larger than
those from $1/N$ effects. We have investigated this by
studying the large $N$ extrapolation of the plaquette,
by calculating the spectrum of the Wilson-Dirac operator and
its hermitian counterpart, and by calculating large Wilson
loops in order to see if we can extract the heavy-quark
potential.

Work along similar lines has recently been
reported in Ref.~\cite{CGU}. These authors simulate
the theory with two adjoint Wilson fermions
on a $2^4$ lattice, and use $N=2-6$. They report
evidence that, for $b=2$, there is a region of quark
masses around the putative critical value, 
including quarks of masses $\sim 1/a$, for which the
center symmetry is unbroken.

This paper is organized as follows. 
In the following section we describe the AEK model and 
the properties of the large-volume theory to which it
would be equivalent were reduction to hold.
In Sec.~\ref{sec:alg} we describe the
algorithm that we use, and show some results concerning its
performance.
Section~\ref{sec:phase_diagram} is the core of the paper, 
in which we use our numerical results to determine
the phase diagram of the AEK model.
We then, in Sec.~\ref{sec:observables}, present
first results for ``observables''---the spectra of the 
Wilson-Dirac operator and its hermitian counterpart,
and large Wilson loops. 
We close in Sec.~\ref{sec:outlook} with a
summary and a discussion of the outlook for future work.
An appendix describes models for eigenvalues of
the Wilson-Dirac operator that are used in Sec.~\ref{sec:observables}.

\section{The AEK model and its putative large-volume equivalent}
\label{sec:model}

The partition function of the single-site theory is
\begin{equation}
Z_{\rm AEK}=\int \prod_\mu DU_\mu  D\psi D\bar\psi 
\,\exp{\left( S_{\rm gauge} + 
\sum_{j=1}^{N_f} \bar \psi_j \, D_{\rm W} \, \psi_j\right)}\,,
\label{eq:Z_AEK}
\end{equation}
where the four $U_\mu$ are SU(N) matrices, while
$\bar\psi_j$ and $\psi_j$
are Grassmann Dirac variables of flavor $j$, living in the
adjoint representation of $SU(N)$.
We use the Wilson gauge action
\begin{equation}
S_{\rm gauge}=2 N \,b\, \sum_{\mu<\nu} {\rm Re}\Tr U_{\mu\nu}^{\rm plaq} \,,
\label{eq:S_gauge}
\end{equation}
where $U_{\mu\nu}^{\rm plaq}$ is the product of links 
around the plaquette in the $\mu,\nu$ plane,
and $b$ is the inverse 't Hooft coupling,
\begin{equation}
b \equiv \frac{1}{g^2N} \,.
\end{equation}
We also use Wilson's lattice Dirac operator 
\begin{equation}
D_W =
1 - \kappa \left[\sum_{\mu=1}^4 \left( 1 - \gamma_\mu\right) 
U^{\rm adj}_\mu
+ \left(1 + \gamma_\mu\right)U^{\dag {\rm adj}}_{\mu}\right]\,,
\label{eq:D_W}
\end{equation}
where $U^{\rm adj}_\mu$ is the adjoint representative of
$U_\mu$, and $\kappa$ is the usual hopping parameter,
related to the bare quark mass by
\begin{equation}
m_0 = \frac{1}{2\kappa}-4\,.
\label{eq:m_0}
\end{equation}
Periodic boundary conditions on both gauge and fermion fields 
have been built into the form of $D_W$.
Throughout this paper we set $N_f=2$.

The theory has a $Z_N^4$ center symmetry, under which
\begin{equation}
U_\mu \longrightarrow z^{n_\mu} U_\mu
\,,
\end{equation}
where $z=\exp(2i\pi/N)$ and $0\le n_\mu < N$ are integers.
Note that $U_\mu^{\rm adj}$ is invariant under this transformation,
so that the fermion action is also invariant.
There is also the single-site version of the gauge symmetry
\begin{equation}
U_\mu \longrightarrow \Omega U_\mu \Omega^\dagger\qquad
[\Omega\in SU(N)]\,.
\end{equation}
Finally, there is an $SO(4)$ flavor symmetry, most easily
seen by writing the action in terms of four 
Majorana fields.\footnote{%
In the continuum this symmetry becomes an $SU(4)$ symmetry, 
but, with our choice of fermion discretization, 
only its $SO(4)$ subgroup remains an exact symmetry of \Eq{eq:Z_AEK}).}

If reduction holds, this single-site theory is equivalent,
when $N\to\infty$, to a theory that has any number of 
lattice sites $L_\mu$ in each of the periodic directions $\mu=1,2,3,4$, 
including the case of $L_\mu=\infty$. The action of the $L_\mu > 1$ 
theories has the same form except that  $U$,
$\psi$, and $\bar\psi$, are now fields having a site index,
$S_{\rm gauge}$ contains a sum over the position of the plaquettes,
and $D_W$ connects fermion fields at adjacent sites. We stress that 
an important feature of reduction is that it relates the single-site
and infinite-volume theories having the same bare parameters,
$b$ and $\kappa$.

We recall some important properties of the infinite-volume theory,
since these will be inherited by the single-site theory if
reduction holds. First, the theory is asymptotically free---$N_f>11/4$
fermions are required to change the sign of the first coefficient
of the \mbox{$\beta$-function}.
Second, although the bare quark mass vanishes when $\kappa=1/8$,
this critical value of $\kappa$ is additively renormalized
because Wilson fermions do not preserve chiral symmetry.
The critical value is shifted to $\kappa_c(b)>1/8$, and
the physical quark mass becomes
\begin{equation}
m_{\rm phys} = \frac1a \left(\frac1{2\kappa}-\frac1{2\kappa_c}\right)\,.
\end{equation}
Here we have introduced the lattice spacing, $a$,
which can be determined, in principle, by fixing
the value for a physical scale, such as a particle mass.
Since the theory is asymptotically free at short distances,
one approaches the continuum limit ($a\to0$) by sending 
$b\to\infty$, and in this limit $\kappa_c(b)\to 1/8$.

The nature of this critical line depends on the
infrared behavior of the theory.
One possibility is that the theory lies below the
conformal window, so that chiral symmetry is spontaneously broken,
much as in QCD.
Then, for $\kappa$ near $\kappa_c$,
one can study the long-distance
behavior and vacuum structure of the lattice
theory using chiral perturbation theory (ChPT).
In particular, close to the continuum limit, one can
use a modified ChPT which includes discretization 
effects~\cite{SharpeSingleton}.
For adjoint fermions, the symmetry breaking pattern differs from that
in QCD, and is $SU(4)\to SO(4)$. 
The required generalization of the analysis of 
Ref.~\cite{SharpeSingleton}
has been given in Ref.~\cite{DelDebbio08}.
One finds that, as in QCD, there are two possible scenarios:
either there is a first-order transition line, at which
the degenerate pseudo-Goldstone ``pions'' attain their
minimal, non-zero mass,
or there are two second-order lines, along which the pions
are massless, and between which there is an Aoki-phase~\cite{Aoki_phase}.
Within the Aoki-phase,
the $SO(4)$ vector symmetry is broken.\footnote{%
We note for completeness that Refs.~\cite{nofirstorder1,nofirstorder2} 
have recently
raised concerns about the consistency of the ``first-order scenario.''}
The width of the Aoki-phase is $\propto a^3$,
and thus shrinks rapidly as one approaches the continuum limit.

A different possibility for the critical line arises if the
massless theory is conformal in the infrared, i.e. if there
is an infrared fixed point.
There is growing numerical evidence that this is the situation
in the $N=2$ theory.
For this theory, the simulations of 
Refs.~\cite{Catterrall:MWT08} and \cite{Kari:MWT08} 
map out parts of the phase diagram.
In particular, there is single, second-order 
transition line emanating from $(b=\infty,\, \kappa=1/8)$, 
while for $b\lesssim 1/4$ the line becomes
a first-order transition. 
(A similar picture holds for an improved fermion action,
but in this case the second-order line extends to 
stronger coupling~\cite{dGSS11}.)
This is not established definitively, and also does not
directly apply to the $N=\infty$ theory that we are interested in.
Nevertheless, this possibility provides a quite different
phase diagram than that which applies when one is outside
the conformal window. One of our aims is to see which
possibility holds at $N=\infty$ (assuming that reduction holds).

\section{Simulation algorithm}
\label{sec:alg}

We simulate the single-site theory using the 
hybrid Monte-Carlo (HMC) algorithm~\cite{HMC}.
Integrating out the fermions leads to 
\begin{equation}
Z_{\rm AEK}=\int \prod_\mu DU_\mu  
e^{S_{\rm gauge}} \det(D_W)^2
\,.
\end{equation}
As usual, $\gamma_5$ hermiticity implies that
$\det(D_W)$ is real, so we can write
\begin{equation}
\det(D_W)^2 = \det(D_W D_W^\dagger) = \det(Q^2)
\,,
\end{equation}
where
$Q = D_W\gamma_5=Q^\dagger$ 
is the Hermitian Wilson-Dirac operator.
Since $Q^2$ has positive eigenvalues, we can represent
its determinant using pseudofermions. Introducing
momenta conjugate to the link variables, we end up
with the HMC Hamiltonian
\begin{equation}
H = \frac12 \sum_\mu \tr(P_\mu^2)
- 2 N b \sum_{\mu<\nu} {\rm Re}{\rm Tr}U_{\mu\nu}^{\rm plaq} 
+ \phi^\dagger Q^{-2}\phi\,.
\label{eq:H_HMC}
\end{equation}
The $P_\mu$ are traceless hermitian $N\times N$ matrices,
while the pseudofermion $\phi$ is complex, lives in
the adjoint representation of $SU(N)$, 
and has an implicit Dirac index. 
It thus has $4 (N^2-1)$ complex components.

In practice, we represent $\phi$ in color space
as a traceless bifundamental, i.e. as a traceless $N\times N$ matrix,
on which $U_\mu^{\rm adj}$ acts as
\begin{equation}
U_\mu^{\rm adj} \phi
\longrightarrow
U_\mu \phi U_\mu^{\dagger}
\,.
\end{equation}
In this way we do not need to explicitly construct 
$U_\mu^{\rm adj}$.\footnote{%
We thank Simon Catterall for stressing this point to us.}
In fact, one could, in principle, keep the trace of $\phi$,
since the singlet field that it represents has no impact on
the dynamics. In particular, one can show that,
in exact arithmetic, the molecular
dynamics (MD) trajectories that are followed are identical with or
without $\tr\phi$ included, as is the change in $H$.
We find, however, that the number of CG iterations required
for a given accuracy is larger if $\tr\phi$ is included, presumably 
because one has to do some work to find the solution for the singlet part.
Thus we always set $\tr\phi=0$.

Our implementation of the HMC algorithm is standard.
We invert $Q^2$ using the conjugate gradient (CG) algorithm, with a
weaker stopping criterion during the MD
evolution than for the accept-reject step.
We require that the residue, $r=b-Q^2 x$, with
$b$ the source,
satisfies $|r|^2/|b^2| < 10^{-5}$ during MD evolution,
finding that any further increase of the cut-off 
leads to a drop in the acceptance.
For the accept-reject step we use $|r|^2/|b^2| < 10^{-15}$,
which makes the error in $\Delta H$ negligible.
Our CG always starts from a vanishing guess, $x_0=0$, which assures
reversibility of the trajectory.
We use trajectories of unit length, and adjust the step size to attain 
acceptances of $0.6-0.85$.

In deriving the gluonic force, one must account for the fact that
each link appears twice in each plaquette. Nevertheless,
the final result has the standard large-volume form:
\begin{equation}
\dot{P}_\mu^U = i N b \sum_{\nu\ne\mu} U_\mu\left[
U_\nu U_\mu^\dagger U_\nu^\dagger
+
U_\nu^\dagger U_\mu U_\nu \right] + h.c.
\label{eq:gaugeforce}
\end{equation}
The fermionic force is
\begin{eqnarray}
\dot{P}_\mu^\phi &=& i \kappa \Bigg\{
(\gamma_5+\gamma_\mu\gamma_5)_{\alpha\beta}
\left[\psi_\beta U_\mu\chi^\dagger_\alpha U_\mu^\dagger
- U_\mu\chi_\alpha^\dagger U_\mu^\dagger \psi_\beta\right]
\nonumber\\
&& - (\gamma_5-\gamma_\mu\gamma_5)_{\alpha\beta}
\left[U_\mu \psi_\beta U_\mu^\dagger\chi^\dagger_\alpha 
- \chi_\alpha^\dagger U_\mu \psi_\beta U_\mu^\dagger \right]
\Bigg\} + h.c.
\label{eq:fermionforce}
\end{eqnarray}
where $\alpha$ and $\beta$ are Dirac indices, 
$\chi = Q^{-2}\phi$, and $\psi=Q \chi$.
Both forces maintain the tracelessness of $P_\mu$.

\begin{figure}[tbp!]
\centerline{
\includegraphics[width=12cm]{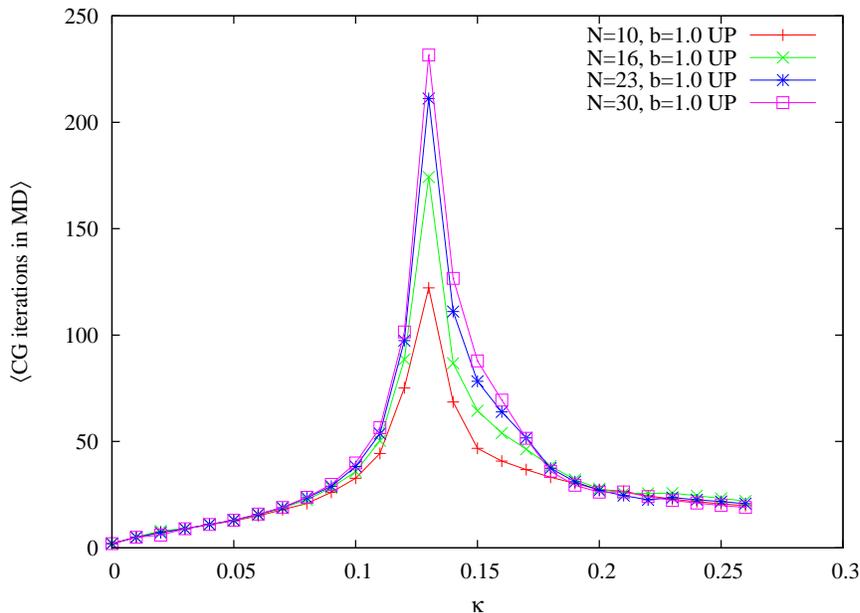}
}
\caption{Average number of CG iterations in the MD updates for various $N$
as a function of $\kappa$ at $b=1.0$. Results are from UP scans.}
\label{fig:CGit_scaling}
\end{figure}

We now discuss the scaling of CPU time with $N$,
which is a key factor in determining how large one can
take $N$.
The core operation---multiplication of $N\times N$ matrices---scales
as $N^3$. The use of the bifundamental
form of $U_\mu$, rather than the adjoint, is crucial here, reducing
the scaling from $N^4$ to $N^3$, as pointed out in Ref.~\cite{CGU}.
The next contribution to the overall scaling comes from the
number of CG iterations, $N_{CG}$. This turns out to depend on the 
proximity to the critical line. 
An example is shown in Fig.~\ref{fig:CGit_scaling},
for the stopping criteria given above.
Away from the critical line, $N_{CG}$ is independent of
$N$, while near the line it grows roughly like $N^{1/2}$.
The third ingredient is the inverse step size, or equivalently the number
of MD steps ($N_{MD}$) per trajectory (for a given acceptance rate). We find
that, to good approximation, this grows linearly with $N$.
Thus, for trajectories of unit length, CPU time
scales as $\sim N^4$ away from the critical line, 
and roughly as $\sim N^{4.5}$ near to the line.
Both scalings are considerable improvements over that for
the Metropolis algorithm used in Ref.~\cite{AEK}, which is
$N^6$ for each SU(2) subgroup update and $N^8$ for an entire update.
On the other hand, our scaling is not as good as the estimate of $N^{3.5}$
given in Ref.~\cite{CGU}, which assumed $N_{MD}\propto N^{1/2}$
and that $N_{CG}$ is independent of $N$, and explicitly
excluded the possible effects of critical slowing down.

We have done both horizontal (fixed $b$)
and vertical (fixed $\kappa$) scans in the $b-\kappa$ plane,
studying the ranges $\kappa=0 -0.6$ and $b=0.05-200$,
although our main focus has been on the smaller ranges 
$\kappa=0-0.26$ and $b=0.35-1.0$.
We use $N=10-30$ in these scans.
Rather than quote a complete set of run parameters we
give a representative example.
We have, for $b=0.35$, $0.5$, $0.75$ and $1.0$, used
27 values of $\kappa$ ($0.0-0.26$ in steps of $0.1$).
At each $\kappa$, we start from the ``configuration''
output from the previous value, thermalize for 500 
trajectories, and then run for 7500 ($N=10$), 5000 ($N=16$),
2000 ($N=23$) or 1000 ($N=30$) trajectories during which
we make measurements every 5 trajectories and store the
configurations every 50.
Each scan is done in both directions---the UP and DOWN scans denoting
increasing or decreasing parameter values (either $\kappa$ or $b$).
To give an example of the CPU time required, the $b=1.0$ UP
scan took 33, 155, 342 and 618 CPU-hours
of a single core on 3.0 GHz Intel Xeon processor,
for $N=10$, $16$, $23$ and $30$, respectively.
Our simulations have been done on local workstations and
using up to 32 CPU cores on a computing cluster. 

We have also done longer runs at several points
in the $b-\kappa$ plane, in which we have gone up to $N=53$.
Details of these runs will be given below.

\section{Phase diagram of the $N_f=2$ AEK model}
\label{sec:phase_diagram}

\begin{figure}[tbp!]
\includegraphics[width=12cm,angle=270]{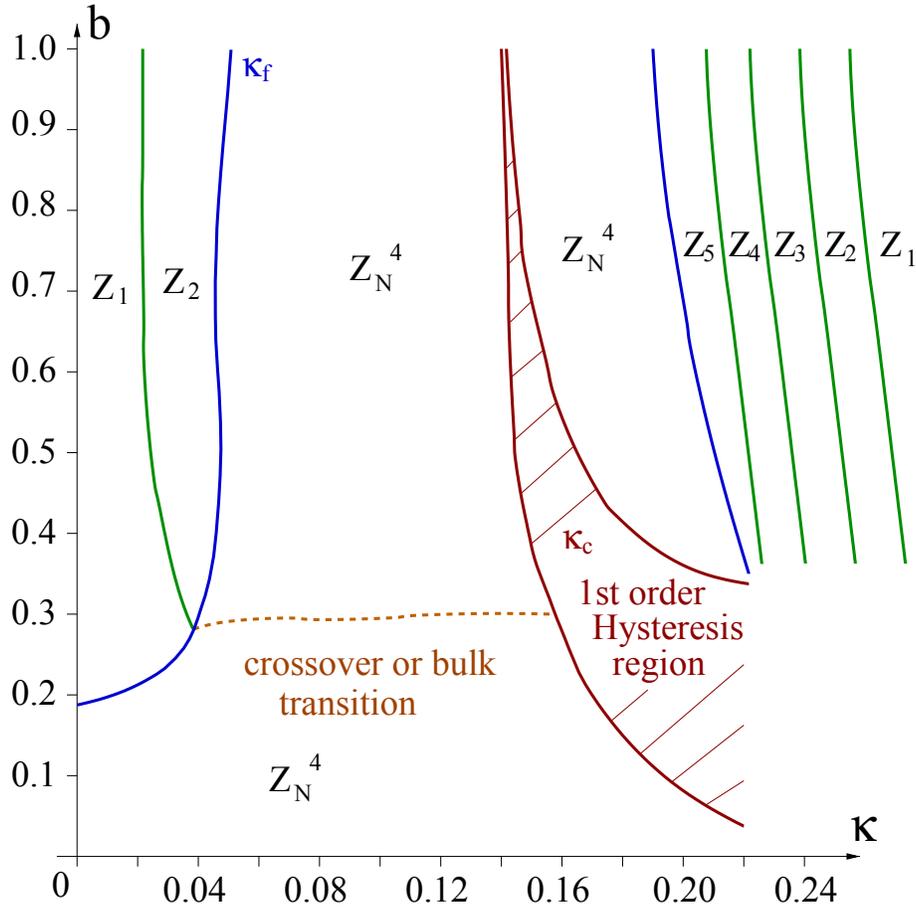}
\caption{Sketch of phase diagram for the $N_f=2$
AEK model in the $\kappa-b$ plane for $N\approx 30$. 
Note that the $\kappa=0$ axis is the EK model.
The positions of phase boundaries are approximate, and depend
somewhat on $N$.
The shaded region at $\kappa_c$ indicates the uncertainty in
the position of what appears to be a first-order transition due
to hysteresis. 
Within each region we note the subset of the $Z_N^4$ center symmetry
that is unbroken, with $Z_1$ indicating complete breakdown. 
The center symmetry is unbroken in the hysteresis region.
The detailed symmetry-breaking
pattern for large $\kappa$ is representative, and depends
to some extent on $N$. For further discussion, see text.
}
\label{fig:phasediagram}
\end{figure}

In this section we present our main results, from which
we deduce the phase diagram sketched in Fig.~\ref{fig:phasediagram}.
The most important conclusion is that there is a ``funnel''
in which the center symmetry is unbroken,
on either side of the first-order transition which we identify
with $\kappa_c$. 
The diagram is qualitatively similar to that found for $N_f=1$~\cite{AEK}.

\subsection{Measured quantities}
\label{subsec:quantities}

To study the gross features of the phase diagram,
we calculate the average of the plaquette, $u_p$, defined by 
\begin{equation}
u_p \equiv \frac1{6N} \sum_{\mu<\nu} Tr(U_{\mu\nu}^{\rm plaq}).
\end{equation}
To study center symmetry breaking, we consider general ``open loops'':
\begin{equation}
K_{n}\equiv \frac1{N}\tr\, U^{n_1}_1\, 
U^{n_2}_2\, U^{n_3}_3\, U^{n_4}_4, 
\quad {\rm with}\ \ n_\mu =0,\pm 1, \pm 2, \dots
\label{eq:Kn}
\end{equation}
where $U^{-n} \equiv U^{\dag n}$. 
These loops transform non-trivially under the center symmetry, 
unless all four $n_\mu$ are integer multiples of $N$.
They are thus order parameters for center-symmetry breaking.\footnote{%
We have used $-5\le n_\mu\le 5$ to keep the quantity of data manageable.
This has the disadvantage that the traces are then insensitive to
symmetry breaking such as $Z_N^4\to Z_{10}$. Histograms
of link eigenvalues, to be discussed below, are, however,
sensitive to such symmetry breaking.}
The simplest choices, on which we focus, are the four Polyakov
loops, $P_\mu = \frac1N\tr U_\mu $ and
the 12 corner variables, $M_{\mu\nu} = \frac1N \tr U_\mu U_\nu$ 
and $M_{\mu,-\nu}=\frac1N\tr U_\mu U^\dag_\nu$,
with $\mu\ne\nu$.

As in the quenched Eguchi-Kawai model, the corner variables turn out
to be particularly useful because they are sensitive to 
partial symmetry breaking~\cite{QEKBS}.
We illustrate this with simple examples.
First, if $U_\mu = \mathbf{1}$ for all $\mu$,
then all the Polyakov loops and corner variables are unity.
This corresponds to complete breaking of the center symmetry.
If instead 
\begin{equation}
\forall \mu: \quad U_\mu = \textrm{diag}\left(
\underbrace{1,\dots,1}_{N/2\ \rm entries},
\underbrace{-1,\dots,-1}_{N/2\ \rm entries} \right)
\end{equation}
(where we have assumed that $N$ is divisible by 4, so that 
$\det U_\mu=1$),
then the Polyakov loops vanish, while all corner variables are unity.
In this case, there is a unbroken  subgroup: 
both $\< P_\mu\>$ and $\<M_{\mu,\pm\nu}\>$ are 
invariant under the $Z_2$ subgroup of $Z_N^4$ generated by
\begin{equation}
\forall \mu:\quad U_\mu \to \pm U_\mu.\label{eq:Z2_subgroup}
\end{equation}
The $U_\mu$ themselves are invariant under 
the combination of \Eq{eq:Z2_subgroup} and a gauge transformation,
the latter being the similarity transformation
which interchanges the first $N/2$ diagonal entries with the second $N/2$ entries.

Such partial symmetry breaking can be discussed in a gauge invariant
way by considering the eigenvalues of link matrices.
For each link we can write
\begin{equation}
U_\mu = W_\mu \Lambda_\mu W_\mu^\dagger\,,
\quad
W_\mu\in SU(N)\,,
\label{eq:linkdecomp}
\end{equation}
with $\Lambda_\mu$ containing the eigenvalues:
\begin{equation}
\Lambda_\mu={\rm diag}\left(
e^{i\theta_\mu^1},e^{i\theta_\mu^2},\dots e^{i\theta_\mu^N}\right)\,.
\end{equation}
Gauge transformations can permute the eigenvalues, but not change
their values.
Center-symmetry transformations change the eigenvalues
by a uniform translation:
$\theta_\mu^a \longrightarrow \theta_\mu^a + 2\pi n_\mu/N$.
Thus a direct way of looking for certain symmetry breaking schemes, 
and understanding their nature, is to look at the distributions of 
the $\theta_\mu$. For example, unbroken center symmetry implies a distribution 
which is invariant under translations by $2 \pi n/N$. 
Partial symmetry breaking occurs when a subgroup of such
translations is unbroken.
In the first example above, the eigenvalues are all clumped,
and all translation symmetries are broken.
In the second example, the eigenvalues form two clumps, and
translation by $\pi$ remains a symmetry.

We use the link eigenvalues in Sec.~\ref{subsec:eigenvalues},
plotting histograms and considering the correlations between
links in different directions.

\subsection{Scans at moderate coupling ($b\le 1$)}
\label{subsec:scans}

In this section we use scans of the plaquette, Polyakov loops and
corner variables to map out the gross features of the phase diagram.

\begin{figure}[tbp!]
\includegraphics[width=12cm]{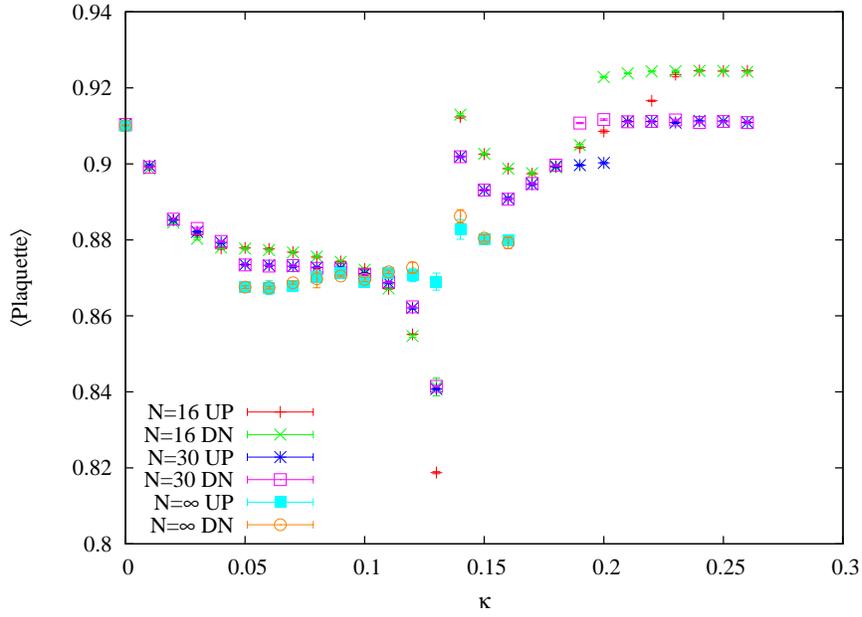}
\caption{Scans (both UP and DOWN) of the average plaquette at $b=1.0$
for $N=16$ and $30$. The results of an
extrapolation to $N=\infty$ (described in the text) are shown in the
central region.
}
\label{fig:scan_b1.0}
\end{figure}

\begin{figure}[tbp!]
\includegraphics[width=12cm]{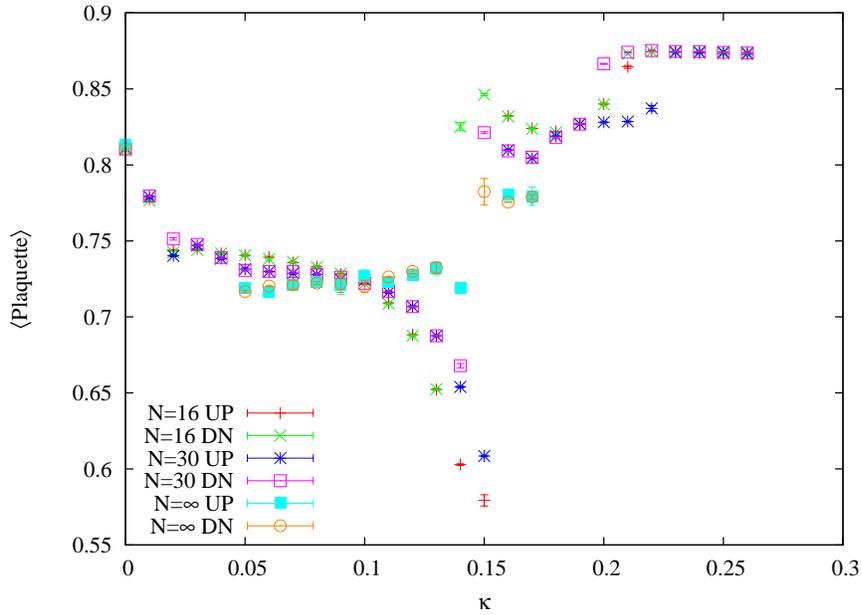}
\caption{As for Fig.~\ref{fig:scan_b1.0} but for $b=0.5$. 
}
\label{fig:scan_b0.5}
\end{figure}

The most interesting values of $b$ are roughly $0.35-1.0$;
this was the range studied in the $N_f=1$ model~\cite{AEK}.
For $N=3$, this corresponds to $\beta=6/g^2=6.3-18$, a range
running from couplings similar to those used in large-volume simulations
to very weak coupling. We have made detailed scans at $b=0.35$, $0.5$,
$0.75$ and $1.0$. That at $b=0.35$ shows a great deal of structure that
is hard to analyze (including large hysteresis and the influence
of a bulk transition), while that at $b=0.75$ interpolates between 
the results at $b=0.5$ and $1.0$.
Thus we show, in Figs.~\ref{fig:scan_b1.0} and \ref{fig:scan_b0.5}
respectively, scans of the plaquette at $b=1.0$ and $0.5$.
We have simulated with $N=10$, $16$, $23$ and $30$, but, for
the sake of clarity, show results only for $N=16$ and $30$.
We also show, in the central region, an approximate estimate of the result at
$N=\infty$, obtained by fitting results at the four values of $N$ 
(or more values, if available) to $c_0+c_1/N+c_2/N^2$.
Such fits will be discussed in Sec.~\ref{subsec:P_M_scaling}.

The results at $b=1$ show three main features:
(i) a change in slope at $\kappa\approx 0.02$
(and possibly another at $\kappa\approx 0.05$),
(ii) a jump at $\kappa=0.13-0.14$,
and (iii) a transition region at $\kappa\approx 0.2$.
These correspond on the phase diagram of Fig.~\ref{fig:phasediagram} to 
(i) the transition region from center-symmetry broken phases
to the unbroken central region,
(ii) the transition line at $\kappa_c\approx 0.13$,
and (iii) the transition to the region of multiple broken phases
for large $\kappa$.
We focus first on the central feature,
presenting our evidence concerning symmetry-breaking below.
An important issue is whether the jump at finite $N$
survives as a first-order transition at $N=\infty$.
Our extrapolations suggest that it does:
although the jump in the plaquette
decreases with $N$, it appears to remain finite at $N=\infty$. 

The conclusion of a first-order transition
is clearer in the results at $b=0.5$.
These show the same qualitative features as for $b=1$, but
the jump in the plaquette is larger, and there is hysteresis
for $N=16$ and $30$.\footnote{%
We note in passing a peculiar phenomenon we have seen for
smaller values of $N$ in the range we consider.
At $N=10$, the UP scans for $b=0.5$ show, in the hysteresis region,
points  having average plaquettes with  non-vanishing imaginary parts.
This breaks the charge conjugation symmetry of the theory, and
is reminiscent of results found in the twisted EK model~\cite{TEKfailure}.
It is because of these points that we do not have an extrapolated
result for the UP scans at $\kappa=0.15-0.17$ in Fig.~\ref{fig:scan_b0.5}.
We suspect that this occurs only in metastable phases,
and view it as a sign of the complicated 
vacuum structure of the single-site theory.}

\begin{figure}[tbp!]
\includegraphics[width=12cm]{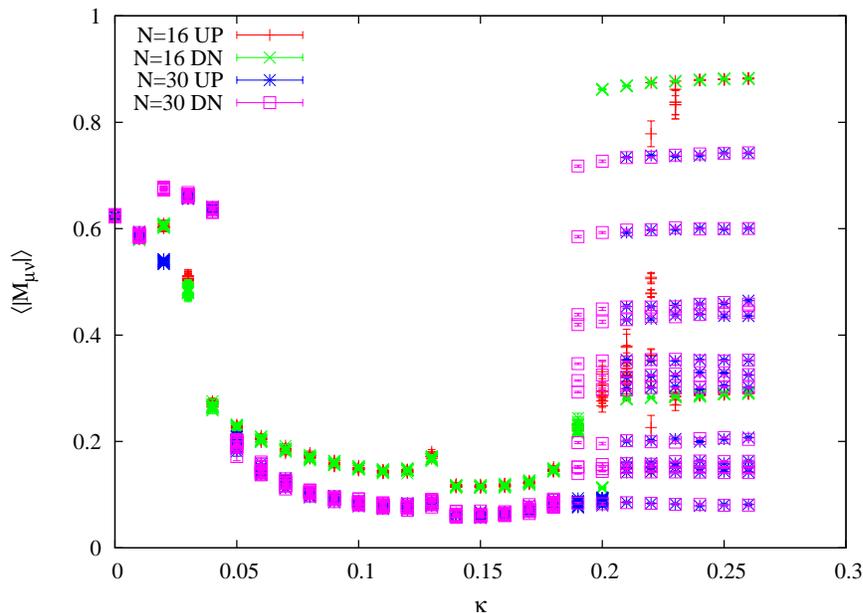}
\caption{Scans of the absolute values of corner
variables for $b=1.0$, for $N=16$ and $30$. 
Results for the 12 independent $|M_{\mu\nu}|$'s
are shown separately. 
}
\label{fig:scanM_b1.0}
\end{figure}

To study center symmetry breaking, we first use scans of
the absolute values of Polyakov loops and the corner variables.
Both should vanish as $N\to\infty$ if the symmetry is unbroken.
The corner variables are more informative and we show an example,
for $b=1.0$, in Fig.~\ref{fig:scanM_b1.0}.
Results for $0.35 \lesssim b < 1.0$ are qualitatively similar.

For both small and large $\kappa$, $\kappa \lesssim 0.05$ 
and $\kappa\gtrsim 0.19$, the corner variables indicate
that the center symmetry is broken.
The nature of this breaking is clarified by the Polyakov loops, $|P_\mu|$,
whose plots we do not show for the sake of brevity.
For $\kappa \lesssim 0.02$ we find $|P_\mu|$ to be non-vanishing
as $N\to\infty$, indicating that the center symmetry is
completely broken. This is
the $Z_1$ phase shown in Fig.~\ref{fig:phasediagram}.
For $0.02\lesssim \kappa\lesssim0.05$ and $\kappa=0.19-0.20$, however, 
$|P_\mu|$ are consistent with zero at $N=\infty$, 
indicating only partial symmetry breaking.
The nature of this partial breaking can be elucidated using
the distributions of
$P_\mu$ and $M_{\mu\nu}$ in the complex plane, 
and using histograms of link eigenvalues.
Some examples of the latter will be shown in Sec.~\ref{subsec:eigenvalues}.

A key issue for reduction is the realization of 
center symmetry in the central funnel,
$0.05 \lesssim \kappa \lesssim 0.18$.
For the values of $N$ used in the scans,
we find no indication of symmetry breaking.
Our evidence is as follows.
Scatter plots of $\langle M_{\mu\nu}\rangle$ and
$\langle P_\mu\rangle$ in the complex plane
show a single distribution centered around the origin,
with averages consistent with zero.
Similarly, the higher-order traces, $K_n$,
which we have evaluated at several positions inside the
funnel, are all consistent with zero.
Finally, histograms of link eigenvalues,
examples of which are shown in Sec.~\ref{subsec:eigenvalues},
are also consistent with the absence of symmetry breaking.

The other key question is whether the funnel remains of
finite width as $N\to\infty$. 
We can see from Fig.~\ref{fig:scanM_b1.0} that the funnel 
narrows with increasing $N$.
In particular, the lower edge of the funnel, which we call $\kappa_f$,
increases from $\kappa_f\approx 0.03$ at $N=10$ 
to $\kappa_f\approx 0.05$ at $N=30$.
To study this question further requires larger
values of $N$, and we defer consideration until
Sec.~\ref{subsec:funnel_scaling}.

\begin{figure}[btp!]
\includegraphics[width=12cm]{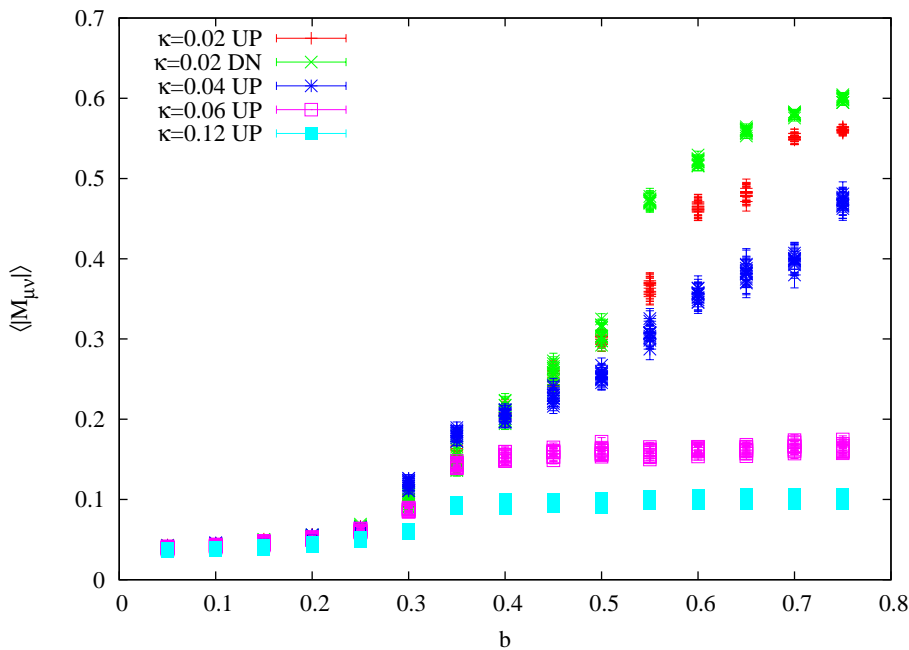}
\caption{
Vertical scans of the absolute values of corner variables
with $N=23$ and for $\kappa=0.02$, $0.04$, $0.06$ and $0.12$.
For $\kappa=0.02$ both UP and DOWN runs are shown.
}
\label{fig:vertM_lowk}
\end{figure}

To complete the study of the phase diagram we have done
several vertical scans with $N=16$, $23$ and $30$.
We show an example of the results in Fig.~\ref{fig:vertM_lowk},
which displays the $\langle|M_{\mu\nu}|\rangle$ for $N=23$
and $\kappa < \kappa_c$. 
For all $\kappa$, there is no indication of symmetry breaking
at strong coupling, $b\lesssim 0.3$, just as for the EK model.
There are possible transitions, however, as we increase $b$.
For example, at $\kappa=0.02$ we see two transitions: 
one at $b\approx 0.3$ and a second at $b\approx 0.55$. 
The first is from a center
symmetric phase to one in which both Polyakov loops and
corner variables are
non-vanishing, consistent with complete symmetry breakdown.
The second is to a phase with large
corner variables and smaller Polyakov loops, which
we interpret as a partially broken phase. 
This is the same ``$Z_2$'' phase apparent in
Fig.~\ref{fig:scanM_b1.0} for $\kappa=0.02-0.04$.

For all other $\kappa$ there is no hysteresis, so we
show only UP scans. At $\kappa=0.04$, we see only a single
transition, at $b\approx 0.3$, and this is directly to
a partially broken phase with only $\langle|M_{\mu\nu}|\rangle$
non-zero. For higher $\kappa$, however,
the symmetry is unbroken on both sides of the jump
at $b\approx 0.3$, and we interpret this as a bulk
transition. It is unclear, however, whether this corresponds
to a phase transition or a crossover as $N\to\infty$.

Vertical runs at higher values of $\kappa$ fill in gaps
left by the horizontal scans, and are part of the input
which leads to the phase diagram of Fig.~\ref{fig:phasediagram}.
For the sake of brevity, however, we do not show any plots here.

\subsection{Scaling of the plaquette, $\langle|P_\mu|^2\rangle$, 
$\langle|M_{\mu\nu}|^2\rangle$}
\label{subsec:P_M_scaling}

In order to study the key question of whether the symmetry unbroken
funnel remains as $N\to\infty$, we have 
extended the calculations to larger values of $N$ at several
several values of $b$ and $\kappa$.

We begin by looking at the average plaquette.
If reduction holds, then, away from $\kappa_c$, the single-site theory
is equivalent at large $N$ to a large-volume lattice theory with
quarks having physical masses of $O(1/a)$.
The long-distance physics of such a theory is that of a pure-gauge
theory with action modified from the pure Wilson form by fermionic effects.
If $\kappa$ is much smaller than $\kappa_c$, these modifications should
be small, since they are proportional to powers of the small 
quantity $\kappa$
(as can be seen using the hopping parameter expansion).
The large-volume theory is thus close to the 
pure-gauge theory with Wilson action.
We can therefore make the semi-quantitative prediction that, 
near the lower boundary of the funnel, $\kappa_f$,
the average plaquette should lie close to the large-volume,
pure-gauge (Wilson action)
value, but depart from that value as one approaches $\kappa_c$.
On the other side of the transition, however,
we do not expect the plaquette to tend to this same value as 
$\kappa-\kappa_c$ increases. This is because $\kappa$ is now
larger, so the action differs 
more significantly from the pure-gauge Wilson form.

\begin{figure}[tbp!]
\includegraphics[width=12cm]{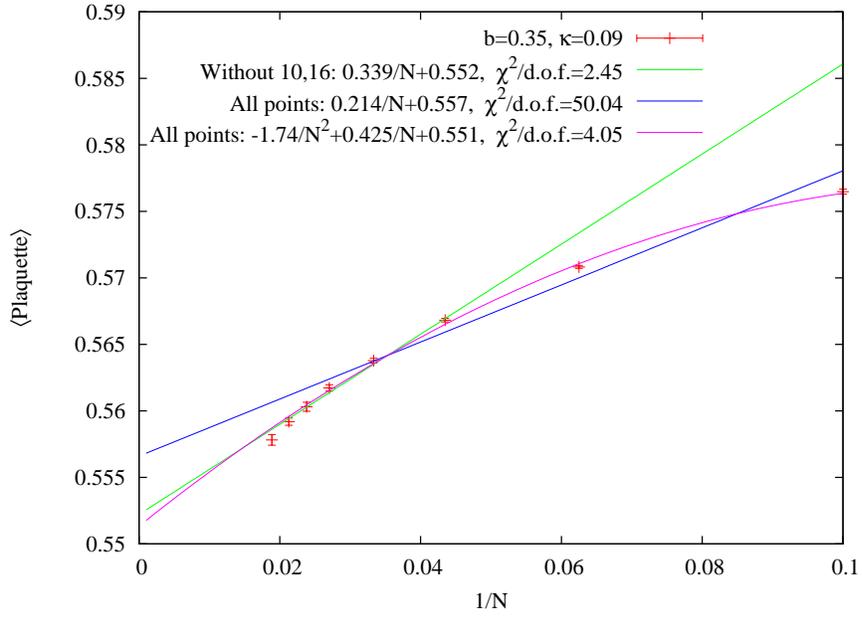}
\caption{Plaquette vs. $1/N$ at $b=0.35$, $\kappa=0.09$, showing
various fits.}
\label{fig:plaq_b35k09}
\end{figure}

\begin{figure}[tbp!]
\includegraphics[width=12cm]{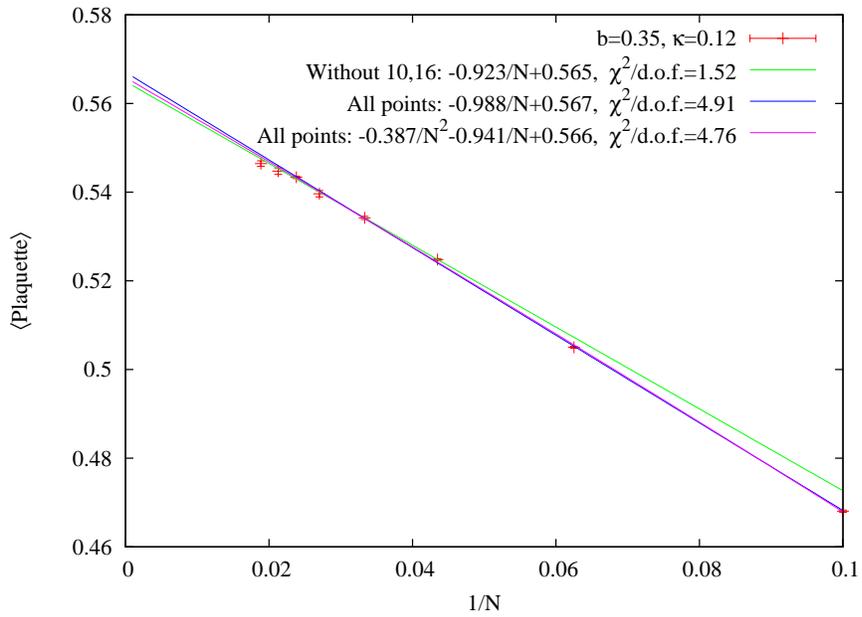}
\caption{As in Fig.~\ref{fig:plaq_b35k09} but for $\kappa=0.12$.}
\label{fig:plaq_b35k12}
\end{figure}

Figures~\ref{fig:scan_b1.0} and \ref{fig:scan_b0.5} 
show that the plaquette has considerable
dependence on $N$, with the slope of this dependence varying with
$\kappa$. We show in Fig.~\ref{fig:plaq_b35k09} an
example of an extrapolation in which the plaquette decreases
with $N$, and in Fig.~\ref{fig:plaq_b35k12} an example in which it
increases. Results are for $N=10-53$ and are plotted versus $1/N$.
We use this variable because we have found that we cannot obtain a
reasonable fit without taking the leading correction to
be proportional to $1/N$ (rather than $1/N^2$). Examples
of fits to $c_0 + c_1/N$ and $c_0 + c_1/N + c_2/N^2$ are shown
in the figures, fitting either to all the data or 
dropping the two lowest values of $N$.
We find that fits of $c_0+c_1/N$ to the highest six values of $N$
are tolerable (probability $p\gtrsim 0.04$) 
for all choices $b$ and $\kappa$ that we have
considered, and use such fits to obtain the
results given in Table~\ref{tab:plaq}.

\begin{table}[tb]
\setlength{\tabcolsep}{3.5mm}
\begin{tabular}{ccccccc}
\hline\hline
$b$ & $\kappa$ & $\chi^2/d.o.f.$ & $c_1$ & $c_0$ & pure-gauge value 
\\ \hline
0.35 & 0.06 & 1.8 & 0.75(4) & 0.549(1) & 0.550 
\\
0.35 & 0.09 & 2.4 & 0.34(4) & 0.552(1) & 0.550 
\\
0.35 & 0.12 & 1.5 & -0.92(3) & 0.565(1) & 0.550 
\\ \hline
1.0  & 0.06 & 0.2 & 0.120(3) & 0.8694(1) & 0.8692 
\\
1.0  & 0.09 & 1.1 & 0.076(3) & 0.8697(1) & 0.8692 
\\
1.0  & 0.12 & 0.6 & -0.248(4) & 0.8709(1) & 0.8692
\\
1.0  & 0.15 & 2.3 & 0.39(1) & 0.8795(4) & 0.8692
\\ \hline
\end{tabular}
\caption{Results from large-$N$ extrapolation of
plaquette expectation values. Extrapolations are done using a fit
of $c_0+c_1/N$ to results at $N=23$, $30$, $37$, $42$, $47$ and $53$.
Results for $c_0$, $c_1$ and $\chi^2/d.o.f.$ are given,
with errors being statistical.
Systematic errors (from different choices of fit function) are
a few times larger than the statistical errors.
Our best estimate of the pure-gauge large-volume expectation value is also 
quoted. The $b=1$ value is obtained from Ref.~\cite{AEK},
while that at $b=0.35$ is obtained from the $N=8$
pure gauge result at $b=0.3504$ from Ref.~\cite{Alltonetal}.
}
\label{tab:plaq}
\end{table}

The results for the plaquette at $N=\infty$ are in striking
agreement with the semi-quantitative prediction explained above.
In particular, for $\kappa=0.06$ and $0.09$ they are consistent
with the pure-gauge large-volume results, while for
$\kappa=0.12$ (close to $\kappa_c$) they begin to differ.
Thus, looking back at the scans of the plaquette in
Figs.~\ref{fig:scan_b1.0} and \ref{fig:scan_b0.5},
we see that the extrapolation to $N=\infty$ leads to
an almost constant value
between the onset of the funnel at $\kappa_f\approx 0.05$
and $\kappa_c$, with the value being close to that of the pure-gauge
theory. 

We also have results for a single point above $\kappa_c$:
$\kappa=0.15$ at $b=1.0$.
We find here that the extrapolated plaquette differs, with
high significance, from that below $\kappa_c$.
This is consistent with our semi-quantitative prediction,
and also indicates that the first-order transition
at $\kappa_c$ survives the large $N$ limit. 

We now return to our numerical finding that the leading
corrections to the plaquette scale as $1/N$. 
This result has also been found in the numerical
studies of Ref.~\cite{AHUY}.
It differs from the naive expectation that, 
with fields in the adjoint irrep,
corrections should be powers of $1/N^2$. We find, however,
that fits to $c_0+c_2/N^2 +c_4/N^4$ are only possible with
very large coefficients ($|c_2|\sim 10-20$, $|c_4|\sim 2500$) 
of alternating signs. We consider these fits unreasonable
since we expect coefficients of $O(1)$.

In fact, there are (at least) two possible sources of $1/N$ terms.
The first can be seen from the perturbative result for
the plaquette when one has a center-symmetric vacuum~\cite{Okawa}
\begin{equation}
u_p = 1 - \frac{1}{8b}(1-1/N) + O(1/b^2)\,,
\label{eq:one_loop_plaq}
\end{equation}
which manifestly contains a $1/N$ correction.\footnote{%
We thank Ari Hietanen for reminding us of this result.}
This correction arises from the fact that in
the decomposition (\ref{eq:linkdecomp}) non-trivial fluctuations
in $W_\mu$ lie in $SU(N)/U(1)^{N-1}$.
In other words, the fluctuations must be off-diagonal, leading
to the factor $N(N-1)=N^2(1-1/N)$.
We expect that the one-loop form (\ref{eq:one_loop_plaq}) should
work reasonably well at $b=1$ (as it does for $c_0$ in
Table~\ref{tab:plaq}) and that the predicted $1/N$ correction
should be most applicable for the smallest values of $\kappa$ 
(where fermionic contributions to the plaquette are minimized).
Indeed the result for $c_1$ at $b=1$, $\kappa=0.06$ lies close
to the prediction of $1/8$.

A second source for $1/N$ corrections are
the ``would-be zero modes'' of the Wilson-Dirac operator $D_W$,
which we discuss in more detail in Sec.~\ref{subsec:spectDW}.
There are $4(N-1)$ of these (corresponding, as in the
gauge case above, to the diagonal generators
of SU(N) in perturbation theory), and they form an
$O(1/N)$ fraction of the total number of modes.
Unless the contribution of these modes is exactly canceled 
by an $O(1/N)$ contribution from the remaining $4(N^2-N)$ eigenvalues, 
these modes can cause observables to depend on odd powers of $1/N$. 
Our results for the
spectrum of $D_W$ suggest that they play an important role
in the dynamics for the values of $b$ at which we simulate.

\bigskip

We now turn to the extrapolations of
$\langle |P_\mu|^2\rangle$ and $\langle|M_{\mu\nu}|^2\rangle$.
In the large-$N$ limit, these can be written, using factorization,
as $|\langle P_\mu\rangle|^2$ and $|\langle M_{\mu\nu}\rangle|^2$, 
respectively, both of which vanish if the center symmetry is unbroken.
Thus an important test of our 
tentative phase diagram is that
$\langle |P_\mu|^2\rangle$ and $\langle|M_{\mu\nu}|^2\rangle$
extrapolate to zero within the funnel.

\begin{figure}[btp!]
\includegraphics[width=12cm]{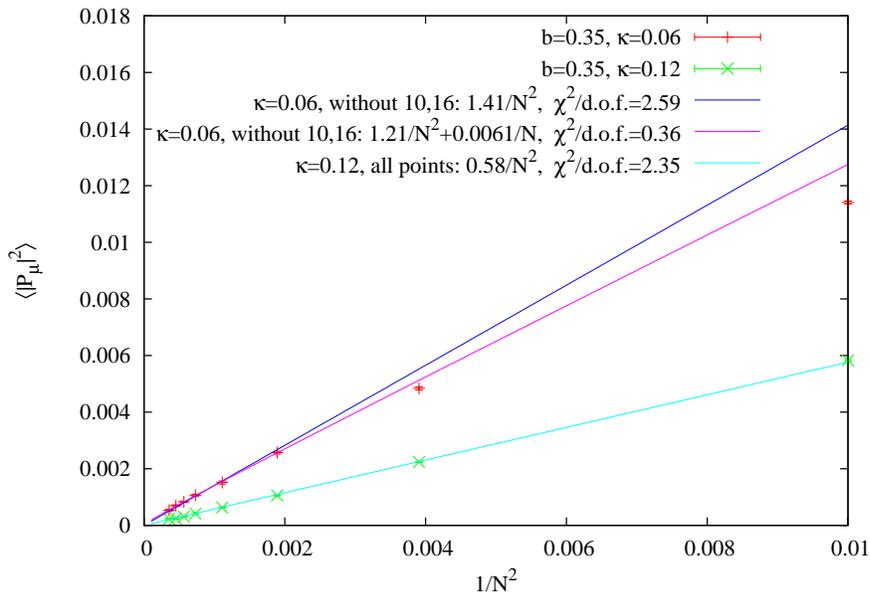}
\caption{$\langle |P_1|^2\rangle$ versus $1/N^2$
for $b=0.35$ and $\kappa=0.06$ and $0.12$,
along with a variety of fits.}
\label{fig:poly_scaling}
\end{figure}

Ordinarily, corrections to factorization are proportional to $1/N^2$,
but, in light of our experience with the plaquette, we might
also see $1/N$ corrections.
In Fig.~\ref{fig:poly_scaling} we plot $\langle |P_1|^2\rangle$
versus $1/N^2$ for $b=0.35$ and two values of $\kappa$.
There is qualitative agreement with a $1/N^2$ fall-off
for both $\kappa$'s, but fits to a pure $1/N^2$ form,
examples of which are shown in the figure, 
have low confidence-levels.
Satisfactory fits (one example of which is shown)
can be found by adding a $1/N$ term and
dropping the lowest two values of $N$.

Results from such fits, for all the values of $\kappa$ and $b$ at
which we have done runs up to $N=53$, are collected in Table~\ref{tab:PandM}.
The fits to $\langle |P_1|^2\rangle$ all have reasonable confidence levels.
The coefficient of the $1/N$ term is small in all cases, 
and in fact is consistent with zero (within $\sim 3\sigma$) 
except for $b=0.35$, $\kappa=0.06$.
We also show results of fits to a constant plus $1/N^2$ term.
The fits are of very similar quality, and the constant turns out
to be very small, and consistent with zero except, again,
at $b=0.35$, $\kappa=0.06$.
We conclude, aside from this one point near to the edge of the funnel,
that the behavior of Polyakov loop is consistent with the
hypothesis that reduction holds in the funnel.

\begin{table}[tb]\footnotesize
\setlength{\tabcolsep}{3.5mm}
\begin{tabular}{ccccccccc}
\hline\hline
Qty&$b$ & $\kappa$ &
$c_1$&$c_2$&$\frac{\chi^2}{\rm d.o.f.}$ &$c'_0$&$c'_2$&$
\frac{\chi'^2}{\rm d.o.f.}$ 
\\ \hline
$\langle |P_{1}|^2\rangle$& 0.35 & 0.06 & 
0.006(1) & 1.21(6) & 0.36 & $9(1)\times 10^{-5}$ & 1.31(1) & 0.39
\\
$\langle |P_{1}|^2\rangle$& 0.35 & 0.09 & 
0.0014(9) & 0.73(3) & 0.67 & $2(1)\times 10^{-5}$& 0.76(2) & 0.73
\\
$\langle |P_{1}|^2\rangle$& 0.35 & 0.12 & 
0.001(3) & 0.57(3) & 0.82 & $0(2)\times 10^{-5}$ & 0.56(1) & 0.84
\\ \hline
$\langle |M_{12}|^2\rangle$& 0.35 & 0.09 & 
0.152(5) & 2.9(1) & 0.42  & 0.0023(3) & 5.3(4) & 2.57
\\
$\langle |M_{12}|^2\rangle$& 0.35 & 0.12 & 
0.036(6) & 3.5(2) & 1.0 & $5(1)\times 10^{-4}$ & 4.1(2) & 1.2
\\ \hline
$\langle |P_{1}|^2\rangle$& 1.0 & 0.06 & 
$-0.0001(3)$ & 1.17(1) & 0.025 & $-1(3)\times 10^{-6}$ & 1.17(1) &
0.025
\\
$\langle |P_{1}|^2\rangle$& 1.0 & 0.09 & 
$-0.0003(4)$ & 0.70(1) & 1.2 & $-5(6)\times 10^{-6}$ & 0.70(1) & 1.2
\\
$\langle |P_{1}|^2\rangle$& 1.0 & 0.12 & 
$-0.0010(3)$ & 0.55(1) & 0.60 & $-1.4(4)\times 10^{-5}$ & 0.54(1) & 0.58
\\
\hline
$\langle |M_{12}|^2\rangle$& 1.0 & 0.06 & 
0.69(3) & -1.6(7) & 0.58 & 0.010(1) & 9(2) & 2.7
\\
$\langle |M_{12}|^2\rangle$& 1.0 & 0.09 & 
0.0053(7) & 6.1(2) & 0.99  & $8(2)\times 10^{-4}$ & 6.9(3) & 1.7
\\
$\langle |M_{12}|^2\rangle$& 1.0 & 0.12 & 
$0.01(1)$ & 5.5(4) & 0.4 & $0(1)\times 10^{-4}$ & 5.4(2) & 0.41
\\
\hline
\end{tabular}
\caption{Results from fits to the large-$N$ behavior of
$\langle |P_{1}|^2\rangle$ and $\langle|M_{12}|^2\rangle$.
Fits are to $N=23$, $30$, $37$, $42$, $47$ and $53$
using $f_1(N)=c_1/N + c_2/N^2$ and $f_2(N)=c'_0 + c'_2/N^2$,
and the quoted errors are statistical.
}
\label{tab:PandM}
\end{table}

\begin{figure}[btp!]
\includegraphics[width=12cm]{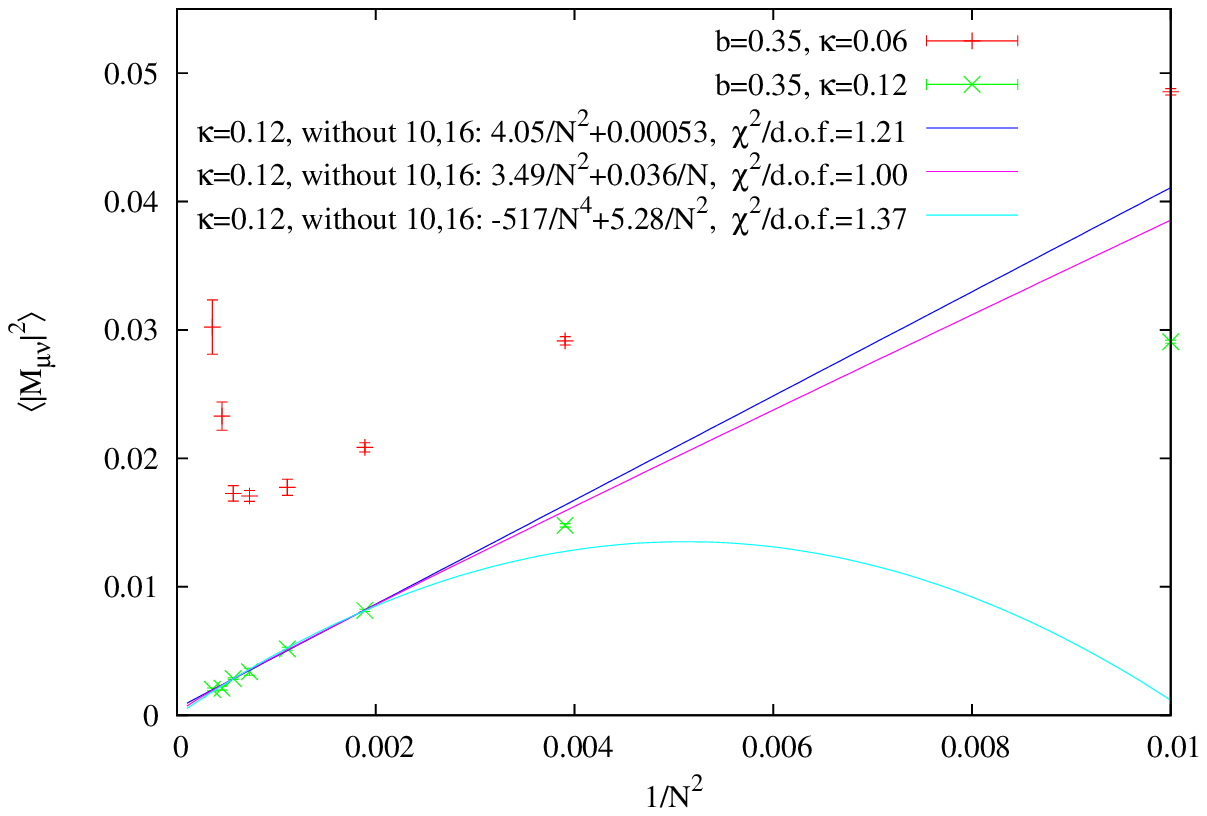}
\caption{As for Fig.~\ref{fig:poly_scaling} but for the
corner variable $|M_{12}|^2$. Only fits to $\kappa=0.12$ are shown.}
\label{fig:corner_scaling}
\end{figure}

Turning to the corner variables, examples of which
are shown in Fig.~\ref{fig:corner_scaling} with full results
collected in the Table,
we find a surprising result: at $b=0.35$, $\kappa=0.06$,
$\langle |M_{12}|^2\rangle$ starts to increase once $N$ exceeds $40$,
and clearly does not extrapolate to zero.
The simplest interpretation of this result is that 
the center symmetry is broken for $N\gtrsim 40$. There is, however,
no other evidence for such symmetry breaking. 
In particular, the distributions of the $M_{\mu\nu}$ and $P_\mu$ 
are approximately uniform around the origin,
the traces $K_{n}$ of Eq.~(\ref{eq:Kn}) are all consistent
with zero, and the link eigenvalues (to be discussed below)
are distributed uniformly.

Instead, our favored interpretation is that the increase in
$\langle |M_{12}|^2\rangle$ with $N$ is due
to the lower edge of the funnel, $\kappa_f$, increasing with $N$.
This increase can be seen (albeit for $b=1$)
in Fig.~\ref{fig:scanM_b1.0} by comparing the results at $N=16$ and $30$.
It is quite possible that, for $b=0.35$,
as $N$ increases, $\kappa_f$ approaches $0.06$,
possibly exceeding this value for $N> 53$. This would lead to the observed
increase in $\langle |M_{12}|^2\rangle$ since this quantity increases
as one approaches the transition (as can be seen in Fig.~\ref{fig:scanM_b1.0}).
This could also explain why our fits to $\langle |P_1|^2\rangle$
were less satisfactory at $b=0.35$, $\kappa=0.06$.

For all the other values of $\kappa$ that we have considered
$\langle |M_{12}|^2\rangle$ decreases monotonically with $N$.
This is exemplified by the
$\kappa=0.12$ results in Fig.~\ref{fig:corner_scaling}.
As for the Polyakov loops, a pure $1/N^2$ fit fails in most
cases, but here we find (cf. Table~\ref{tab:PandM})
that the addition of a $1/N$ term usually
leads to a better fit than the inclusion of a constant.
We also find that, in almost all cases,
 the required $1/N$ (or constant) term has
a coefficient which differs significantly from zero.
We have also done fits to $c_2/N^2+c_4/N^4$
(an example is shown if Fig.~\ref{fig:corner_scaling})
but the fits require very large coefficients having opposite
signs, a fine-tuning which we consider unlikely to be the
correct description.
Overall, we think the most reasonable fits are those to $c_1/N+c_2/N^2$, 
because they have the highest confidence levels, 
and because we have seen in the plaquette that $1/N$ terms are needed.

\bigskip
The results presented so far are consistent with the funnel
(in which center symmetry is unbroken) remaining of finite width
as $N\to \infty$, so that reduction holds for masses up to $O(1/a)$.
We cannot definitively draw this conclusion, however, because of the
following scenario. Imagine that the funnel width vanishes (for any fixed $b$)
as $N\to\infty$. Then, for each $b,\kappa$ point in the putative
funnel, symmetry breaking would occur at a (possibly large, but) finite,
value of $N$. Nevertheless, there would be a finite range of $N$ for which
the symmetry is unbroken, within which the arguments for reduction hold.
Appropriate variables (such as the plaquette) would equal infinite
volume values up to corrections proportional to powers of $1/N$.
Thus, within this range, it might appear that one can extrapolate to the
symmetry-unbroken $N=\infty$ limit, but this would in fact not be the case.
The results for the plaquette at $b=0.35$, $\kappa=0.06$ in 
Fig.~\ref{fig:poly_scaling} are an example of such misleading scaling,
since we have strong evidence from the corner variables that $\kappa_f> 0.06$
at this $b$ for large enough $N$.

\subsection{$N$-scaling of the funnel width}
\label{subsec:funnel_scaling}

In the light of the results in the previous subsections it is important to
directly study the $N$ dependence of the funnel width. We have done so by
focusing on $\kappa_f$, the lower edge of the funnel. If we can show that
$\kappa_f$ remains below $\kappa_c$ as $N\to\infty$, then the funnel remains
open.

\begin{figure}[btp!]
\includegraphics[width=12cm]{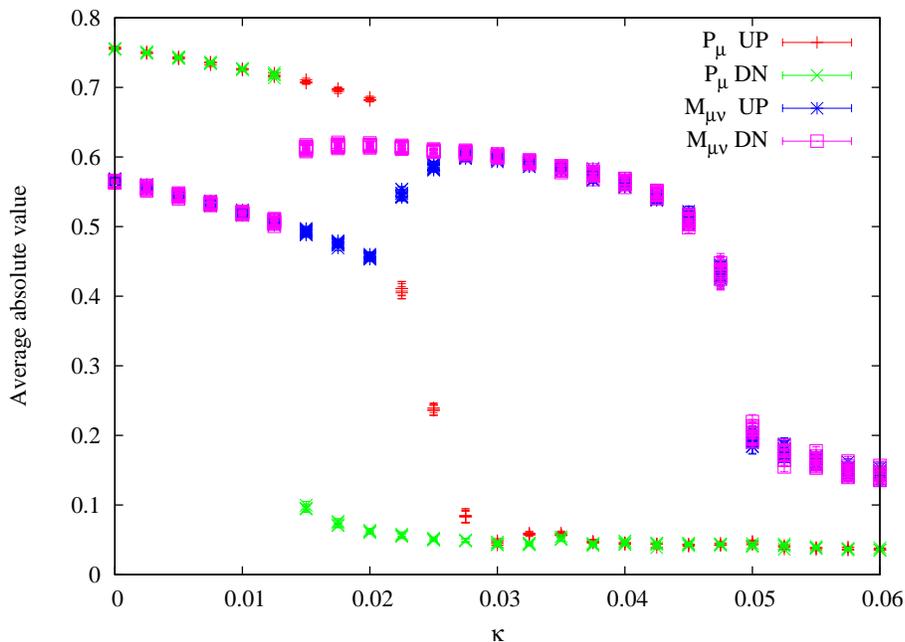}
\caption{Fine scans of absolute values of all Polyakov loops and corner
variables in the low $\kappa$ region, for $N=30$ at $b=0.75$.}
\label{fig:scanM_b0.75_lowkappa}
\end{figure}

To investigate this issue we have done fine scans of the small $\kappa$ region,
an example of which is shown in Fig.~\ref{fig:scanM_b0.75_lowkappa}. There are
two phases before one enters the funnel: a $Z_1$ phase for $0\le \kappa \lesssim
0.02$ and a $Z_2$ phase from $0.02\lesssim \kappa \lesssim 0.05$. The transition
between the first and second phase shows significant hysteresis, while that
between the second phase and the funnel does not. 

\begin{figure}[btp!]
\includegraphics[width=11cm]{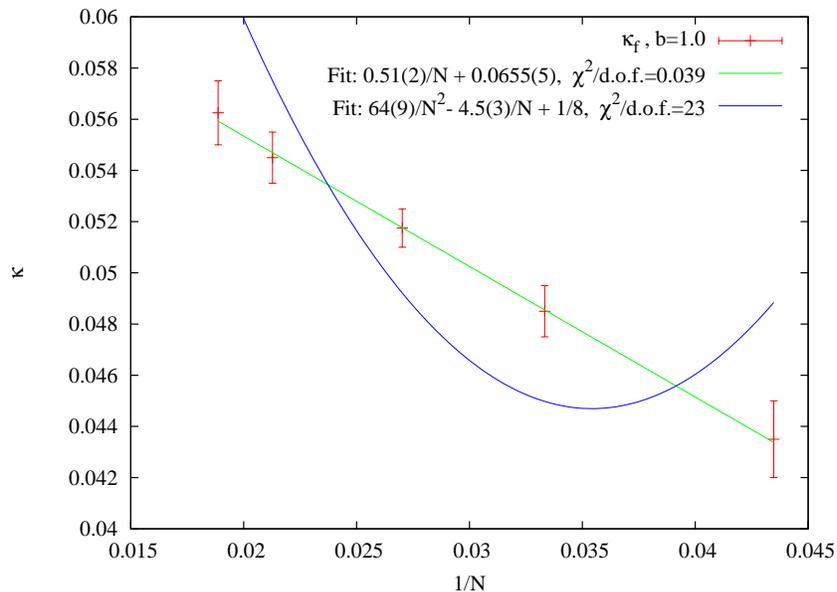}
\caption{The dependence of $\kappa_f$ (lower edge of the funnel) on $1/N$ for
$b=1$. The fit functions are discussed in the text.}
\label{fig:kappa_f_vs_N}
\end{figure}

Determining $\kappa_f(N)$ to high precision 
is a significant numerical challenge. We have thus
focused on a single value of coupling, $b=1$.
For this $b$, Fig.~\ref{fig:scan_b1.0} shows that
$\kappa_c$ lies in the range $0.13-0.14$.
We have done very fine scans near the edge of the funnel (roughly
$\kappa=0.02-0.07$) with $N$ up to 53, and find that the corner variables are
the most useful in determining the transition.
We are able to pin down the transition, conservatively, to
about $\delta \kappa = \pm 0.001$.
The transition from the funnel is to a $Z_2$ phase for most $N$
(as in Fig.~\ref{fig:scanM_b0.75_lowkappa}), but to a
$Z_3$ phase for $N=47$ and $53$.
The results are plotted against $1/N$ in Fig.~\ref{fig:kappa_f_vs_N}, and show
remarkable linearity (note that, as earlier, we have excluded 
$N=10$ and $16$---including them requires adding a $1/N^2$ term 
to obtain a satisfactory fit).
Two fits are shown. The first is to $c_0+c_1/N$, and has a very small
$\chi^2/{\rm d.o.f.}$, with a reasonable coefficient of $1/N$.
It yields $\kappa_f(N=\infty)=0.0655(5)$, a value far below $\kappa_c$.
The second fit is to $\kappa_c + c_1/N + c_2/N^2$, with the intercept fixed to
$\kappa_c=0.125$. This fit is extremely poor, and it gets even worse for
$\kappa_c=0.13-0.14$.
We have also done the corresponding fits to the alternative quantity
$am_f=1/2\kappa_f-1/2\kappa_c$, using $\kappa_c=0.125$ or $0.13$, and find
consistent results.
We conclude that, at least at this value of $b$, the funnel has finite width
when $N=\infty$, so that reduction holds.\footnote{%
Note that, if we use the
linear fit, then the funnel at $b=1$ passes $\kappa=0.06$ when $N\approx 92$.
Thus the successful extrapolations of the plaquette, 
$\langle|P_1|^2\rangle$ and
$\langle |M_{12}|^2\rangle$ for $b=1$, $\kappa=0.06$, presented in
Tables~\ref{tab:plaq} and \ref{tab:PandM}, are examples of the phenomenon
described above in which reduction only holds for a window of values of $N$.}

\subsection{Distributions of link eigenvalues}
\label{subsec:eigenvalues}

We find that histograms of link eigenvalues provide
very useful information on the nature of symmetry breaking on either
side of the funnel. They are also sensitive to patterns of
symmetry breaking in which both Polyakov loops and corner variables vanish,
and thus provide a more stringent test that the symmetry is indeed
unbroken in the funnel.
In this section we present examples of the results that allow us to
fill in the details of the phase diagram of Fig.~\ref{fig:phasediagram}.

\begin{figure}[tbp!]
\subfigure[$\ N=30$, $b=0.35$, $\kappa=0.09$, \mbox{600 configs, all links}]
{\label{fig:hist_N30b35k09}
\includegraphics[width=6.5cm]{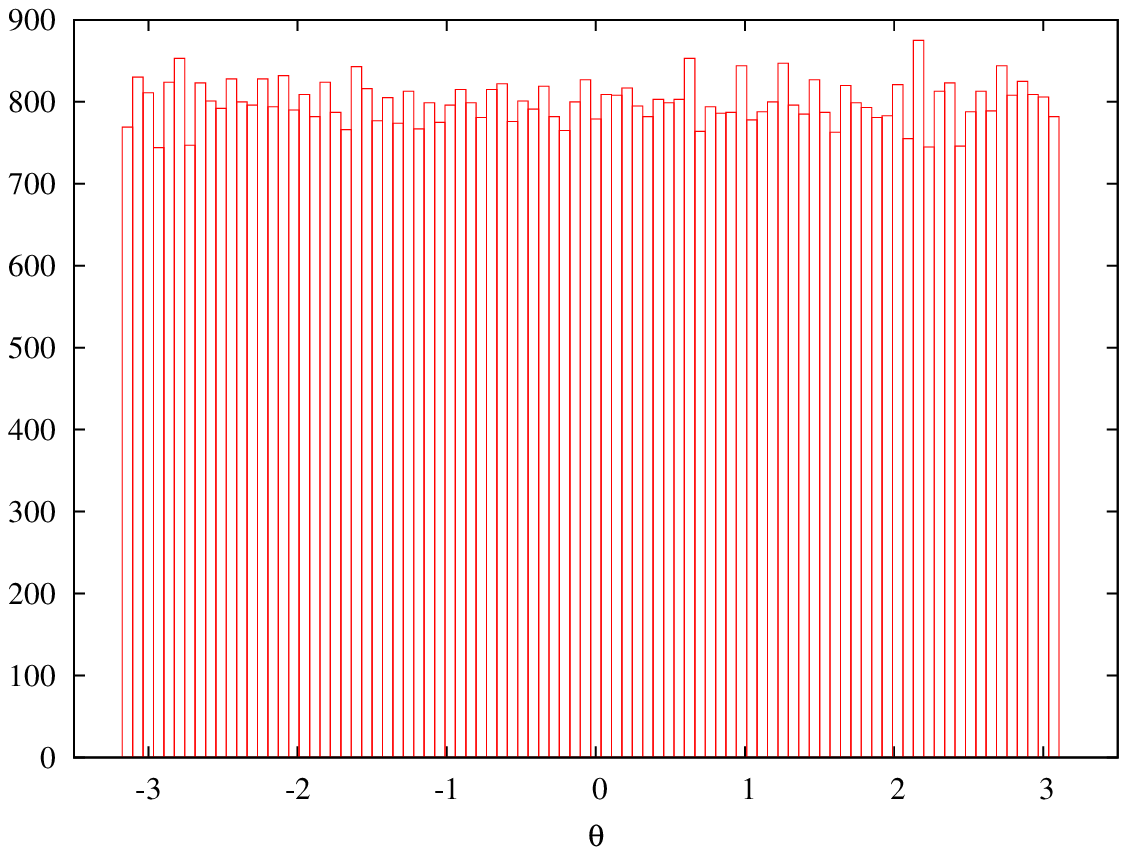}}
\hspace{0.5cm}
\subfigure[$\ N=23$, $b=1.0$, $\kappa=0.01$, \mbox{2000 configs, all links}]
{\label{fig:hist_N23b1k01}
\includegraphics[width=6.5cm]{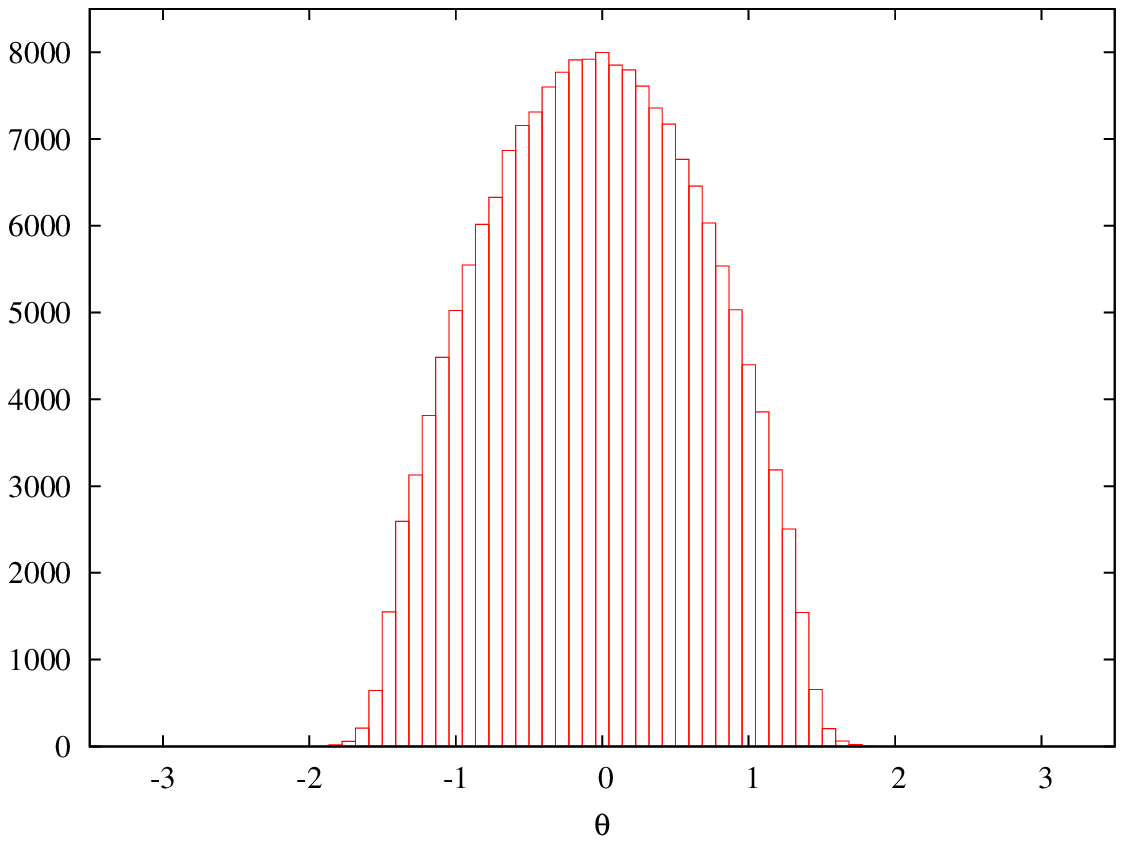}}\\[6pt]
\subfigure[$\ N=23$, $b=1.0$, $\kappa=0.03$, \mbox{2000 configs, $U_2$ only}]
{\label{fig:hist_N23b1k03}
\includegraphics[width=6.5cm]{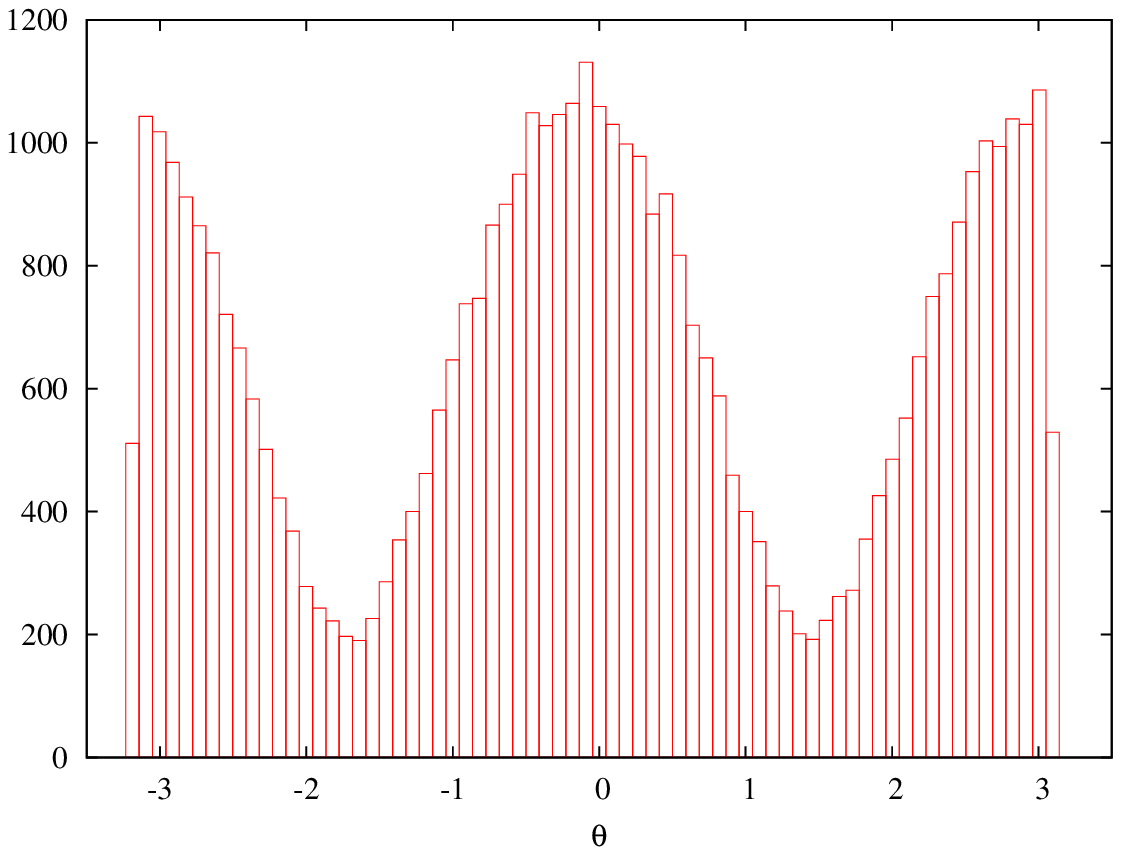}}
\hspace{0.5cm}
\subfigure[$\ N=16$, $b=1.0$, $\kappa=0.24$, \mbox{5000 configs, $U_3$ only}]
{\label{fig:hist_N16b1k24}
\includegraphics[width=6.5cm]{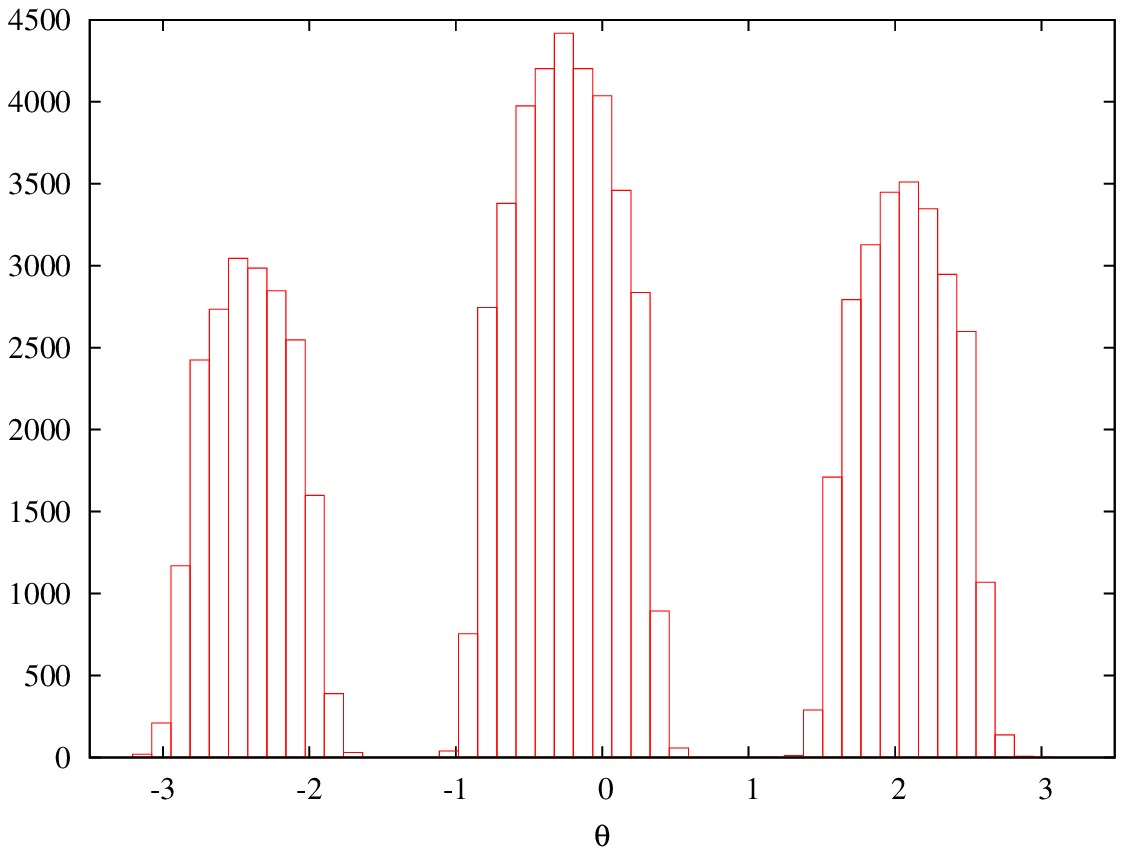}}\\[6pt]
\subfigure[$\ N=16$, $b=0.35$, $\kappa=0.22$, \mbox{3000 configs, $U_3$ only}]
{\label{fig:hist_N16b35k22}
\includegraphics[width=6.5cm]{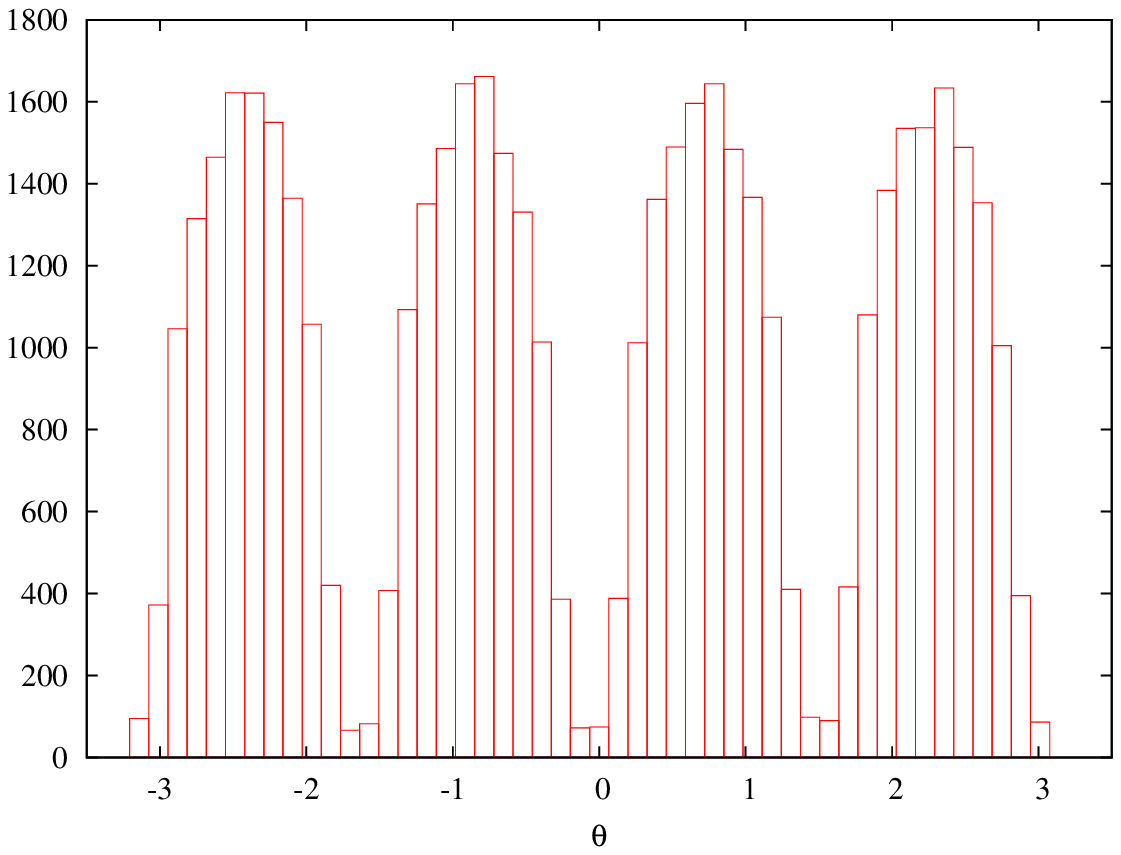}}
\hspace{0.5cm}
\subfigure[$\ N=30$, $b=1.0$, $\kappa=0.23$, \mbox{1000 configs, $U_1$ only}]
{\label{fig:hist_N30b1k23}
\includegraphics[width=6.5cm]{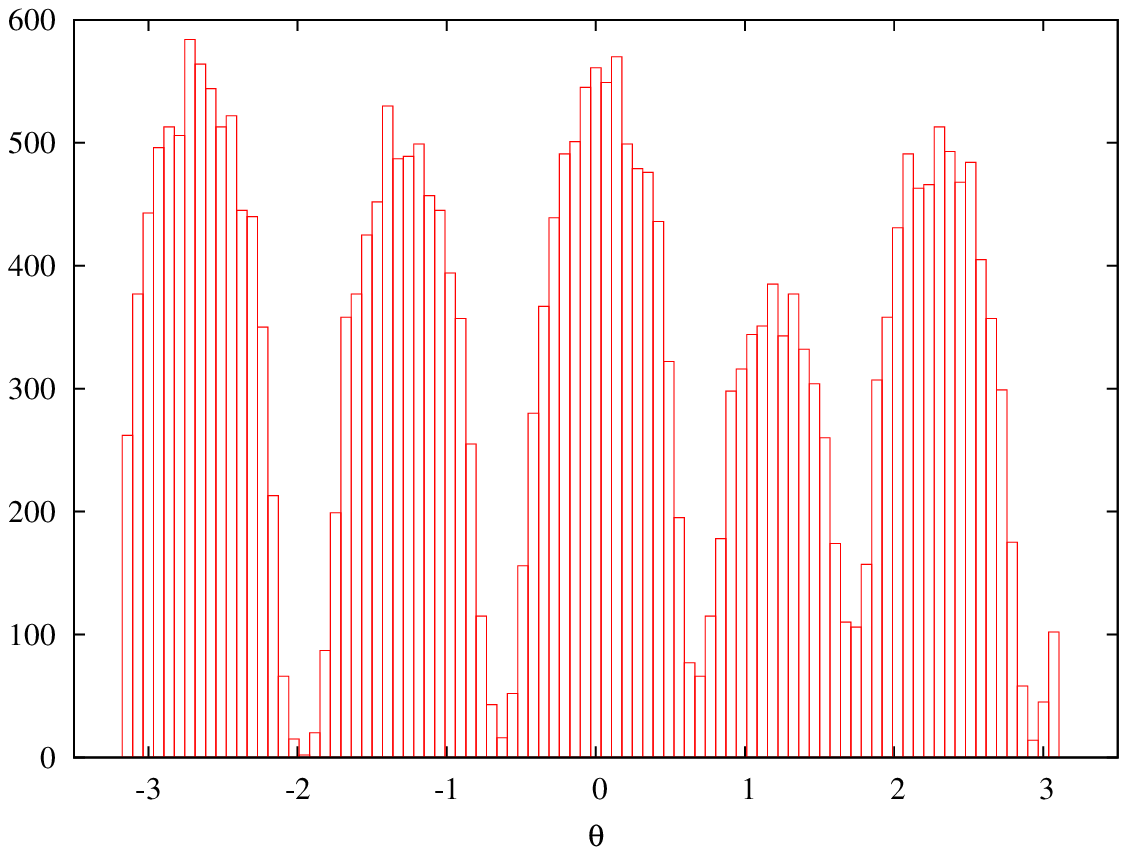}}
\caption{Histograms of the phases $\theta_\mu^a$ of the link eigenvalues. More
details of the binning are discussed in the text.}
\label{fig:hist_leig}
\end{figure}

We begin with an example of a histogram within the funnel,
shown in Fig.~\ref{fig:hist_N30b35k09}. The eigenvalues are taken to have the
range $-\pi<\theta_\mu^a \le \pi$, and are collected in $3N$ bins of width
$2\pi/(3N)$. Thus $Z_N$ symmetry implies invariance under periodic translations
by multiples of 3 bins. In fact the distribution is consistent with being
uniform.\footnote{%
As a check on our code, we have calculated the distribution in
the $b=0$, $\kappa=0$ limit, i.e. for the Haar measure on the links,
and obtain the theoretically expected form, which is $Z_N$-invariant,
but does oscillate within each $Z_N$ segment, although the 
amplitude of the oscillations falls as $1/N$~\cite{NN03}.}

When we move outside the funnel the attraction between eigenvalues leads to
formation of clumps in the complex plane. The number of clumps $k$ identifies
the approximate remnant $Z_k$ symmetry and generally decreases as we move
away from the funnel. In Fig.~\ref{fig:hist_leig} we provide several
examples of clumping patterns. Figs.~\ref{fig:hist_N23b1k01} and
\ref{fig:hist_N23b1k03} present $Z_1$ and $Z_2$ phases in the small $\kappa$
region while Figs.~\ref{fig:hist_N16b1k24}, \ref{fig:hist_N16b35k22} and
\ref{fig:hist_N30b1k23} 
show $Z_3$, $Z_4$ and $Z_5$ phases in the large $\kappa$
region. Note that the partial symmetry breaking can also be seen in the corner
variables giving complex patterns (compare Fig.~\ref{fig:scanM_b1.0}). Polyakov
loops are much less sensitive to this partial symmetry breaking, since they
almost vanish due to the approximate $Z_k$ symmetry.

We find that the remnant symmetry is not always exact. 
For example, in
Fig.~\ref{fig:hist_N23b1k03} we have two clumps for $N=23$ and in
Fig.~\ref{fig:hist_N16b1k24} we have three clumps for $N=16$. Therefore the
eigenvalues cannot be equally distributed between the clumps 
and the symmetry is only approximate. 
Even in Fig.~\ref{fig:hist_N30b1k23}, which shows five clumps
for $N=30$, we see that the clumps are not even and correspond to 7,6,7,4,6
eigenvalues, respectively. We also find that different runs can have different
patterns of eigenvalue clumping, 
e.g. 7,7,6,6,4 versus 7,6,6,6,5, but that it is
rare for the clumping to change during a run. Thus it appears that there are
competing ``vacua'' which are not exactly related by center symmetry
transformations. 

\begin{figure}[btp!]
\includegraphics[width=12cm]{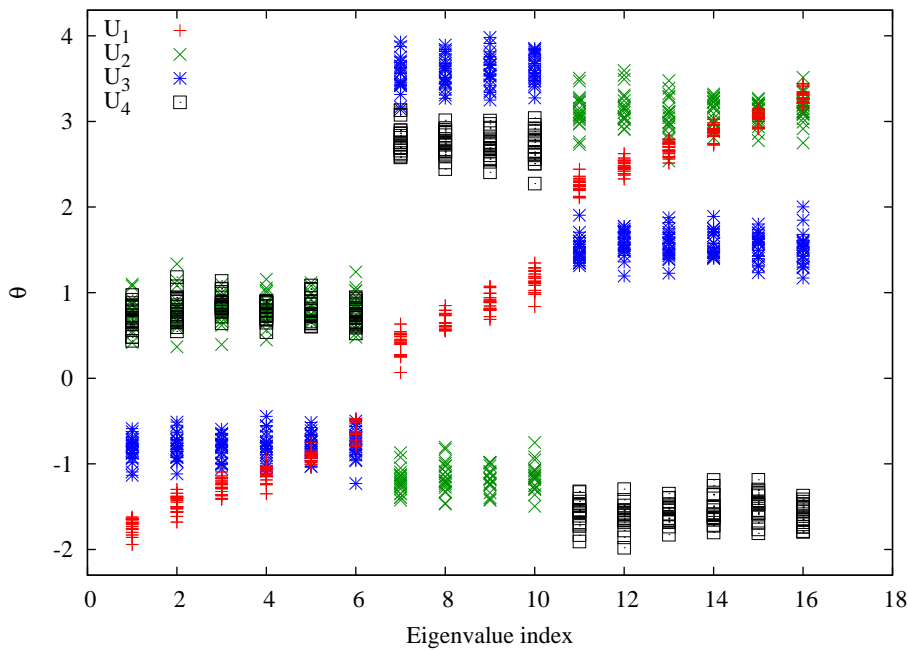}
\caption{Results for the phases of the diagonal elements of $U_\mu$ for 20
thermalized configurations at $N=16$, $b=0.35$ and $\kappa=0.23$ on an UP scan.
Phases are determined in a gauge such that $U_1$ is diagonal with the phases
ordered. For further discussion, see text.}
\label{fig:thetacorrN16b35k23}
\end{figure}

To fully understand the symmetry breaking, we need to know whether there are
correlations between the eigenvalues of different links. What we find is that,
whenever the center symmetry is broken, the eigenvalues for all four links are
highly correlated. To illustrate this, we use a case with three clumps which
makes the results easy to visualize. 
Figure~\ref{fig:thetacorrN16b35k23} shows the resultant 
clumping and correlations. Here we apply a gauge
transformation which diagonalizes $U_1$ and orders the phases $\theta_1^a$, and
then look at the phases of the diagonal elements of $U_{2,3,4}$. These matrices
are close to diagonal, so these phases are presumably close to those of
their eigenvalues. Recall that there is no ambiguity in the ordering of the
diagonal elements once we specify the order for $U_1$. 
The result shows that the
clumps (of 6, 4 and 6), while being positioned at different angles, are almost
completely correlated between all four links,
 and do not change during the Monte-Carlo
evolution. Because of these correlations, the approximate remnant of
the center symmetry for the parameters of Figs.~\ref{fig:hist_N16b1k24}
and \ref{fig:thetacorrN16b35k23} is $Z_3$, and not $Z_3^4$.

Once outside the funnel, the number of clumps decreases as we move to higher
$\kappa$. We have extended some runs to $\kappa=0.6$ and find that the UP scans
end up in a two clump state, while the DOWN runs, which begin from an ordered
start, begin with a single clump, and then have a transition, as $\kappa$ is
decreased, to two (well separated) clumps. In fact, the transition appears to
occur in stages where more and more eigenvalues peel off from the initial clump.
As $\kappa$ is further decreased, the number of clumps increases until we enter
the funnel and there is no longer any clumping. The largest number of clumps
depends on $N$, and the largest we have observed is five, as shown in
Fig.~\ref{fig:hist_N30b1k23}.

The changes in clumping for the large $\kappa$ values, surveyed above, 
are qualitatively consistent with the arguments
presented in Ref.~\cite{AHUY} based on the one-loop free-energy for the link
eigenvalues. Decreasing $\kappa$ from a large value corresponds to reducing the
quark mass $am$. For large $am$, gluonic interactions dominate the free energy,
and lead to attraction, and thus a single clump. As $am$ is reduced, the
fermionic contributions lead to repulsion at large eigenvalue separations
(corresponding to large momenta, so that the mass term is unimportant), while
there remains attraction for small separations.  This allows the possibility of
two clumps. Reducing $am$ further the repulsion becomes important for smaller
eigenvalue separations, and clumps are pulled apart into a greater number of
stable clumps. At the same time, quantum fluctuations within each clump are
always present, so that the clumps have a finite width 
(which is proportional to $b^{-1/4}$ for weak coupling). 
Eventually, as the number of clumps increases,
the distance between the clumps is smaller than the widths, and the clumping is
washed out. 

We would expect a similar sequence of clumpings to occur as we increase 
$\kappa$ from zero, since this also corresponds to reducing $am$. 
This is indeed what we observe, although the maximal number of
clumps is smaller on this side of the funnel.
For $N< 23$ we only find a $Z_1$ phase, for $23\leq N < 47$ we see
both a $Z_1$ and $Z_2$ phase (see Fig.~\ref{fig:scanM_b0.75_lowkappa}),
while for $N=47$ and $53$ we find $Z_1$, $Z_2$ and $Z_3$ phases.
At larger values of $b$, the arguments of Ref.~\cite{AHUY}
imply that the maximal number of clumps should increase.
Indeed, we do find that, as $b$ increases,
the $Z_3$ phase appears at smaller values of $N$.
 
\subsection{Results at large $b$} 
\label{subsec:large_b}

The perturbative calculations of Refs.~\cite{HollowoodMyers2009,BBPT} 
lead us to expect that the center-symmetry-unbroken funnel will
close as $b\to\infty$, so that, in the continuum limit,
reduction only holds for $m=0$ (for $N\to\infty$).
In addition, Ref.~\cite{AHUY} makes a prediction for how rapidly the
funnel should close: its width in $am$ (and thus in $\kappa$)
should be proportional to $b^{-1/4}$. This is because the
width of each clump of link eigenvalues is predicted to
scale to zero proportional to $b^{-1/4}$. A second prediction
(which is explained in the previous subsection)
is that, at the edge of the funnel,
there should be multiple phases with differing numbers of clumps,
and that the maximum number of clumps should
increase with $b$ (as long as $N$ is large enough). 
In other words, a behavior similar to that we have already seen
on the right side of the funnel ($Z_1\to Z_2 \to \dots \to Z_5$;
see Fig.~\ref{fig:phasediagram}) extends to larger groups as $b$ increases.

We have investigated these predictions by doing scans in $\kappa$
for $N=10$ (and, in some cases $N=30$) 
at $b=5$, $10$, $50$ and $200$. We find that the HMC
algorithm mostly performs well even at these very weak couplings.
We did have to reduce the step-size as $b$ increases, such that 
$N_{MD}/{\rm acceptance}$ increases roughly like $\sqrt{b}$.
We also find that, for $b=200$, thermalization for each new value of $\kappa$
sometimes takes longer than our allotted 450 trajectories.
On the other hand, the number of CG iterations gradually
decreases with increasing $b$.
We also note that run histories of observables
show no indication of correlation times that are close to the
number of trajectories we were using for measurements (7500).

\begin{figure}[tbp!]
\includegraphics[width=12cm]{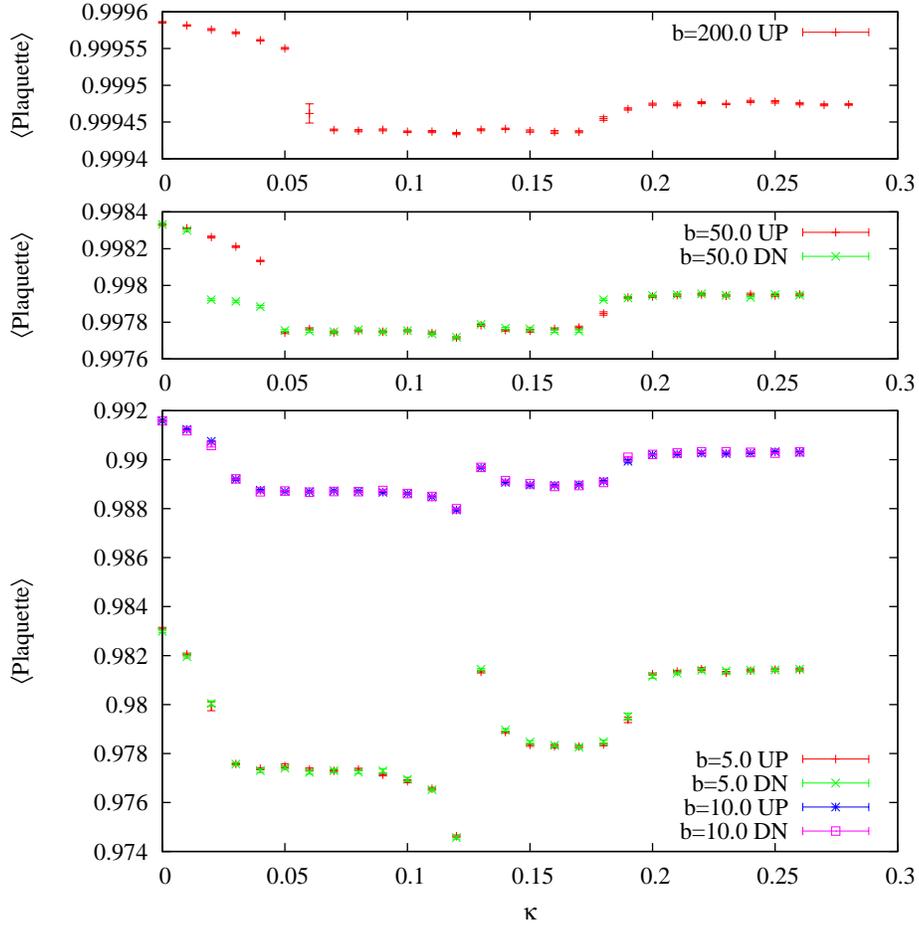}
\caption{The average plaquette in scans at extremely high $b$ for $N=10$.
Note the highly compressed vertical scale at large $b$.}
\label{fig:plaq_highb}
\end{figure}

Results for the plaquette are collected in Fig.~\ref{fig:plaq_highb}.
The general shape of each curve is similar to those at $b=1$
(see Fig.~\ref{fig:scan_b1.0})
but the jump at the putative $\kappa_c$ falls rapidly with
increasing $b$. This is qualitatively consistent with the
expectations from chiral perturbation theory if this is the
first-order scenario of Ref.~\cite{SharpeSingleton}. 

\begin{figure}[btp!]
\includegraphics[width=12cm]{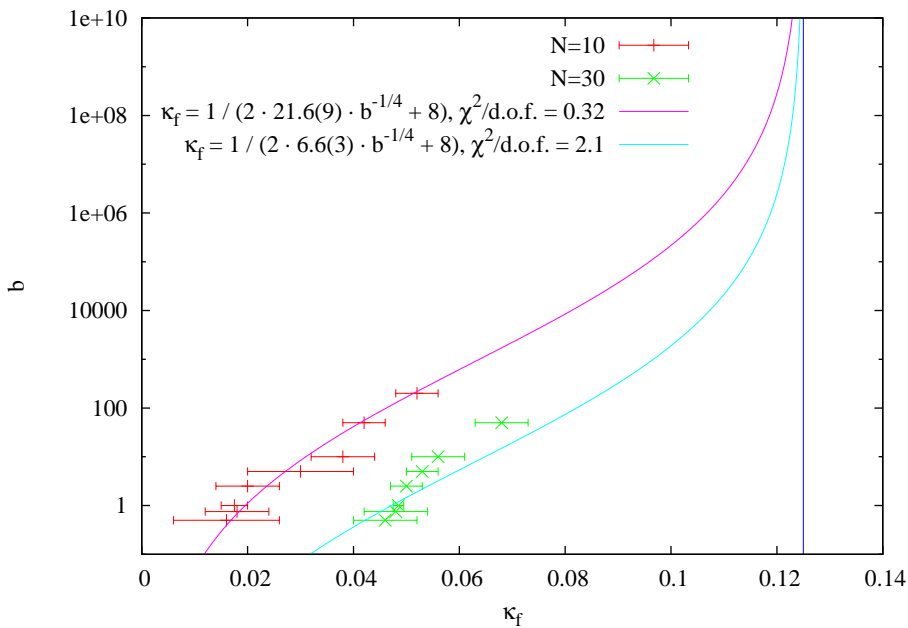}
\caption{The dependence of $\kappa_f$ on $b$ for
$N=10$ and $30$. The vertical (blue) line marks \mbox{$\kappa_c(b=\infty)=1/8$}.
The fit functions are discussed in Sec.~\ref{subsec:large_b}.}
\label{fig:kappa_f_vs_b}
\end{figure}

To verify the hypothesis of Ref.~\cite{AHUY}
that $am_f =1/2\kappa_f-1/2\kappa_c\propto b^{-1/4}$ 
we analyze the lower edge of
the funnel as a function of $b$. It would obviously be advantageous to repeat
the analysis of Sec.~\ref{subsec:funnel_scaling} for all values of $b$;
unfortunately this is numerically too demanding. We do, however, have estimates
of $\kappa_f$ at $N=10$ and $30$ for a wide range of $b$. These are shown in
Fig.~\ref{fig:kappa_f_vs_b}. We find that it is harder to determine
$\kappa_f$ for $b$ away from unity. For smaller values, e.g. $b\approx 0.35$,
the transition becomes smoother. For much larger values, 
there is significant hysteresis (as seen in the middle panel of
Fig.~\ref{fig:plaq_highb}).
The net result is that the errors in $\kappa_f$ are much larger than
those at $b=1$.

Figure~\ref{fig:kappa_f_vs_b} also shows fits to 
$am_f= c\; b^{-1/4}$, with
$\kappa_c$ set to $0.125$ for all values of $b$ for simplicity.
The fit at $N=10$ is good, while that at $N=30$ is poorer.
Better fits at $N=30$ can be obtained using estimates for the
actual value of $\kappa_c$ at each $b$, but these estimates have
sufficient uncertainty that the resulting errors in $m_f$ are 
substantially increased, so that the agreement with the theoretical
form is less significant.
Overall we conclude that our results are consistent with the predicted
dependence on $b$.

The figure also indicates that the narrowing of the funnel as $N$ increases
holds for all values of $b$. 
We do not attempt to extrapolate to $N=\infty$ from $N=10$ and $30$,
as our experience with $b=1$ indicates that $N=10$ is
not in the asymptotic $1/N$ region. However, given that we do find a finite
width when $N\to\infty$ at $b=1$, 
the observation of mild dependence on $b$ suggests that the
funnel will remain of finite width also at other values of $b$.

\section{Measurements inside the ``funnel''}
\label{sec:observables}

In this section we make a detailed study of the funnel region
in which reduction appears to hold using results from $N=10-53$.
We consider in turn
the spectrum of the single-site Wilson operator
$D_W$, the spectrum of $Q^2$, and, finally, attempt to extract a
physical observable---the heavy-quark potential---from large Wilson loops.

\subsection{Spectrum of $D_W$}
\label{subsec:spectDW}

One way of viewing the equivalence of single-site
and large-volume theories is that the space-time volume
is being packaged inside the gauge matrices.
It is thus useful to introduce an effective size, $\Leff$,
and corresponding effective volume, $\Leff^4$, and study
their scaling with $N$. What we mean by $\Leff$ is
that the single-site theory leads to the same physical
results as the theory on an $\Leff^4$ volume with $\Neff$ colors,
up to corrections suppressed by powers of $1/\Neff$.
Clearly there is a trade-off between increasing $\Leff$ and
increasing $\Neff$. Here we take the approach of holding $\Neff$
fixed, but large enough that $1/\Neff$ corrections to quantities
of interest are small, and then asking how $\Leff$ scales with $N$.

Within this framework, the most conservative possibility is
provided by orbifold-based demonstration of volume independence~\cite{KUY2}.
This demonstration also provides an explicit example of the
packaging of the volume into the gauge matrices:
the $N\times N$ link matrices are partitioned
into blocks of size $\Neff\times \Neff$, with $\Neff=N/\Leff^4$,
only $\Leff^4$ of which are non-zero, and the resultant orbifolded theory
is an $SU(\Neff)$ gauge theory on an $\Leff^4$ lattice.
Equivalence is demonstrated by holding $\Leff$ fixed,
and taking $N$, and thus also $\Neff$, to infinity.
Instead, using the approach espoused above, if we hold $\Neff$ fixed,
then increasing $N$ leads to $\Leff\propto N^{1/4}$.

A less conservative possibility is obtained,
following Refs.~\cite{AMNS,U04}, by assuming that all
entries in the link matrices are used in the packaging of
the large volume theory (not just 1 out of every $\Leff^4$ as in
the orbifold construction). 
This leads to $\Leff^4 \sim N^2$ or $\Leff \propto N^{1/2}$.
There is also a more optimistic possibility, $\Leff\propto N$,
which is motivated in the Appendix.

The spectrum of the fermion matrix, $D_W$, can help distinguish
these possibilities, as well as give insight into the nature of
corrections to the large-$N$ limit. 
We expect, if reduction holds, that the spectrum should resemble
that of a large-volume four-dimensional theory
on an $\Leff^4$ lattice. In particular,
for weak couplings, $b\gtrsim 1$, the spectrum
should have the familiar five ``fingers'' which reach down
to the real axis. The number of fingers is a direct
indicator of the dimensionality (there are $d+1$ in $d$ dimensions),
and the distance of the eigenvalues in the fingers from the real axis 
should scale as $1/\Leff$.
These points are discussed in more detail in the Appendix.

We now show some representative results for the spectrum of
$D_W(m_0=0)$ from our simulations. The operator in the determinant is
\begin{equation}
D_W(m_0) = 2\kappa \left[ 4 D_W(0) + \frac{1}{2\kappa} - 4\right]
=\frac{1}{4+m_0} \left[4 D_W(0) + m_0\right]\,,
\end{equation}
so that eigenvalues of $4 D_W(0)$ close to $\lambda=-m_0=4-1/2\kappa$
are suppressed. Since the spectrum is bounded,
$0 \le {\rm Re} \lambda \le 8$,
the determinant suppression is important only for $\kappa>1/8$.
We also note that, unlike on a lattice with an even number
of sites in each direction, 
the spectrum is not symmetric under reflection
about the ${\rm Re}\lambda=4$ axis,
Thus the first (the leftmost) and fifth fingers are not related by symmetry, 
and neither are the second and fourth.
If such a symmetry holds approximately, it indicates the
presence of reduction.

\begin{figure}[tbp!]
\subfigure[$\ \kappa=0.01$]
{\label{fig:dw_N16_b1_k0.01}
\includegraphics[width=7.5cm]{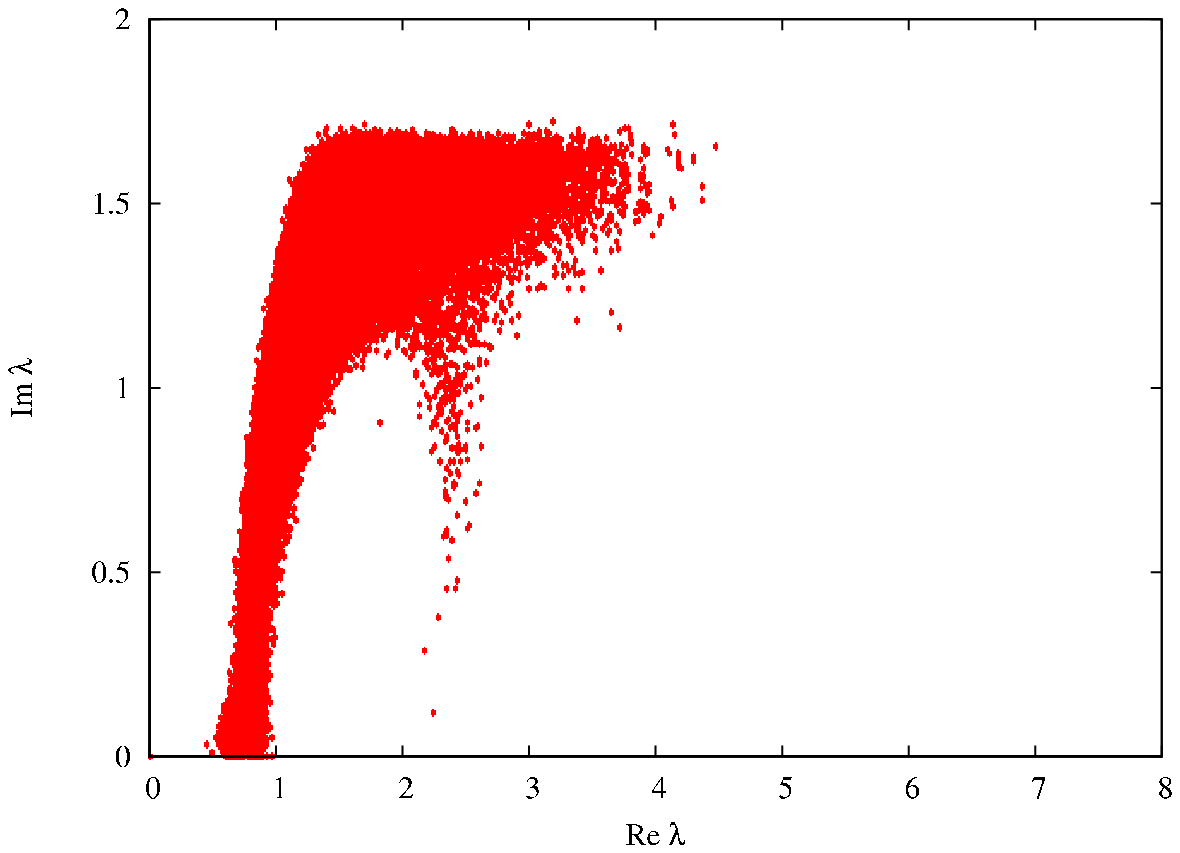}}
\hspace{0.3cm}
\subfigure[$\ \kappa=0.03$]
{\label{fig:dw_N16_b1_k0.03}
\includegraphics[width=7.5cm]{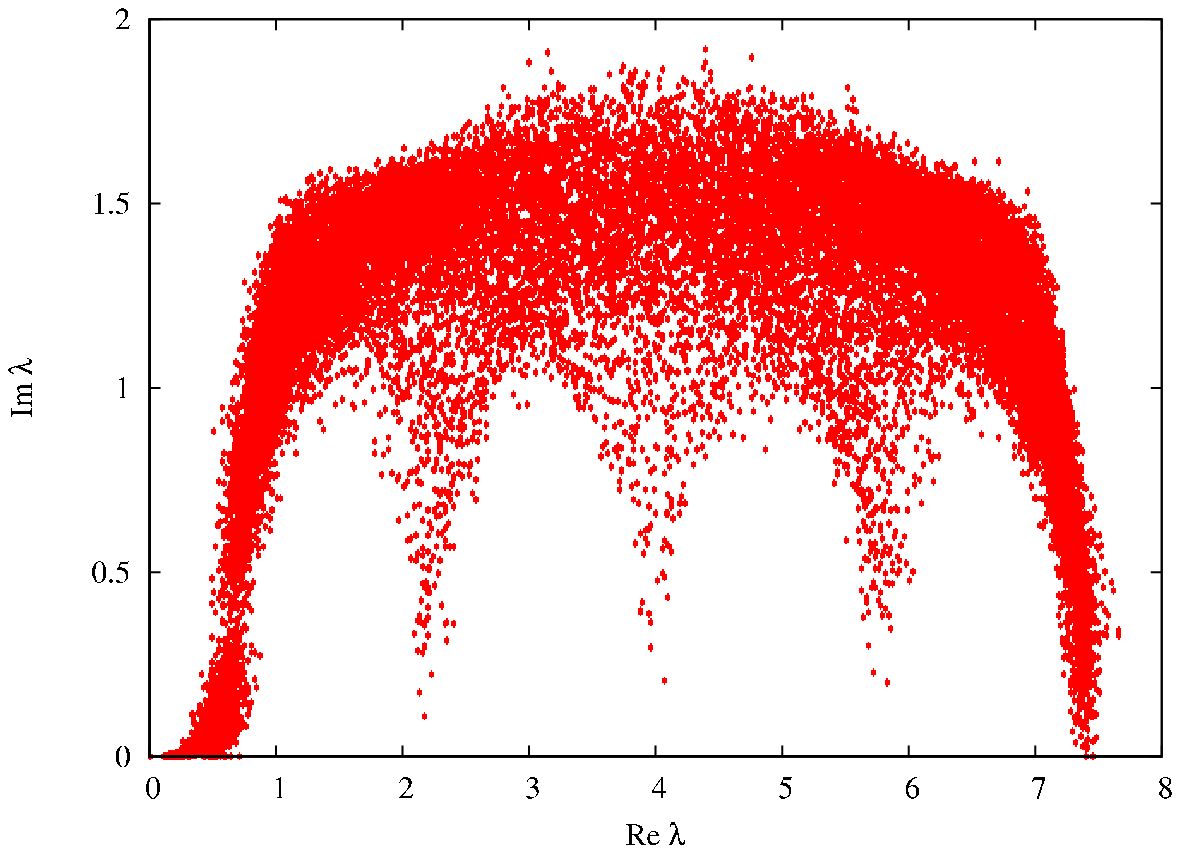}}\\[6pt]
\subfigure[$\ \kappa=0.12$]
{\label{fig:dw_N16_b1_k0.12}
\includegraphics[width=7.5cm]{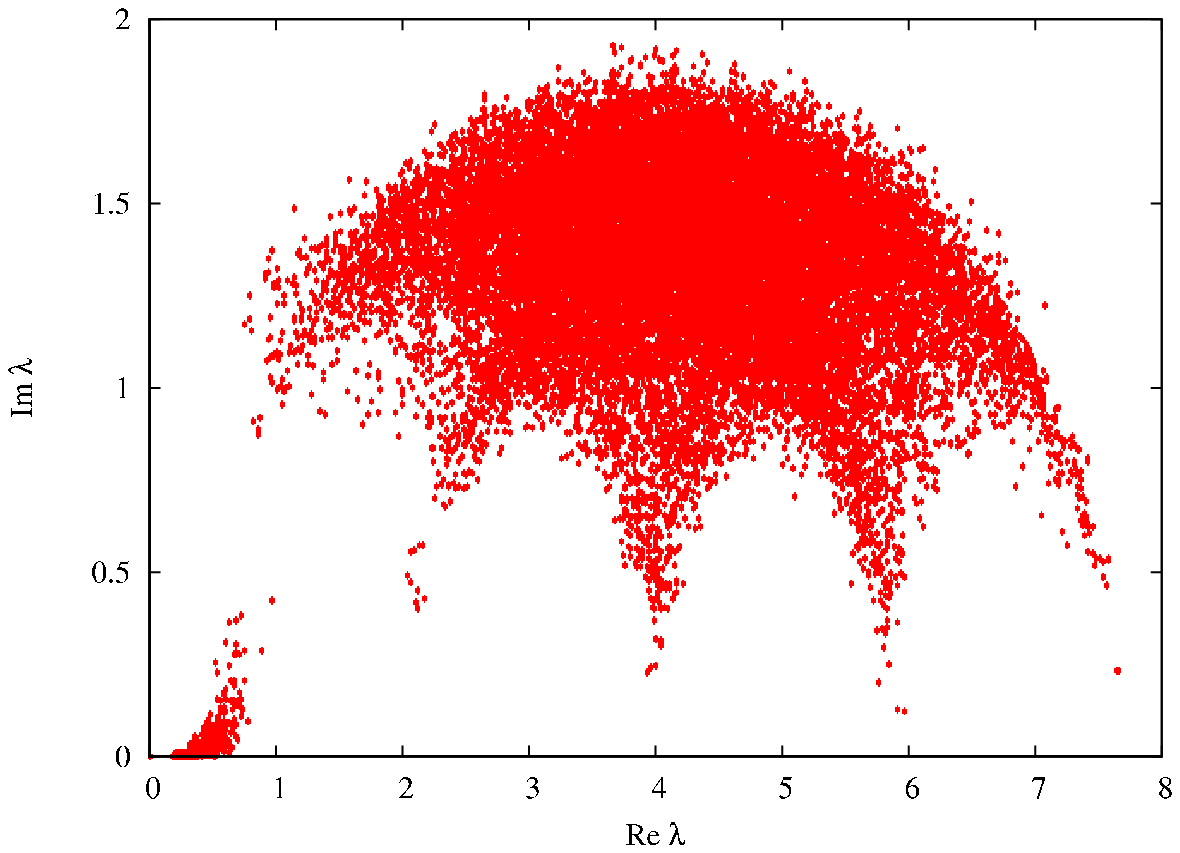}}
\hspace{0.3cm}
\subfigure[$\ \kappa=0.14$]
{\label{fig:dw_N16_b1_k0.14}
\includegraphics[width=7.5cm]{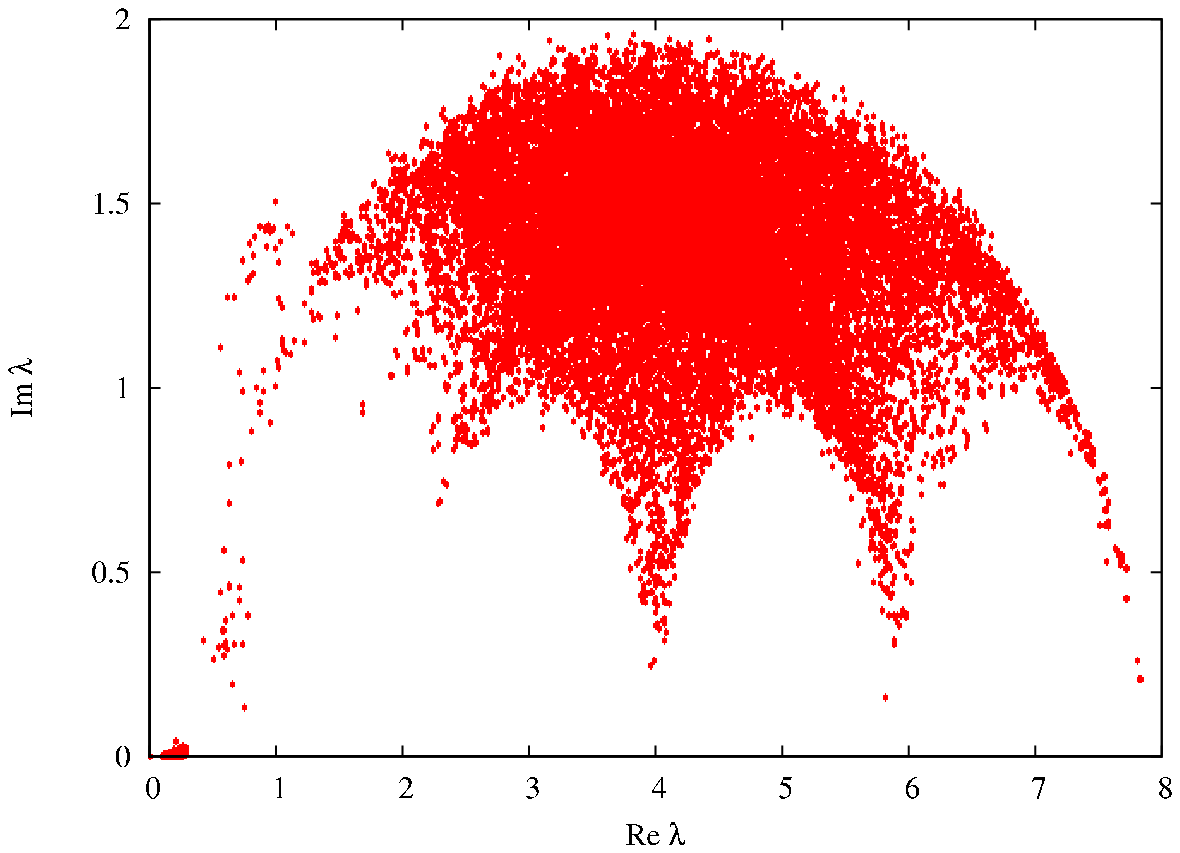}}\\[6pt]
\subfigure[$\ \kappa=0.17$]
{\label{fig:dw_N16_b1_k0.17}
\includegraphics[width=7.5cm]{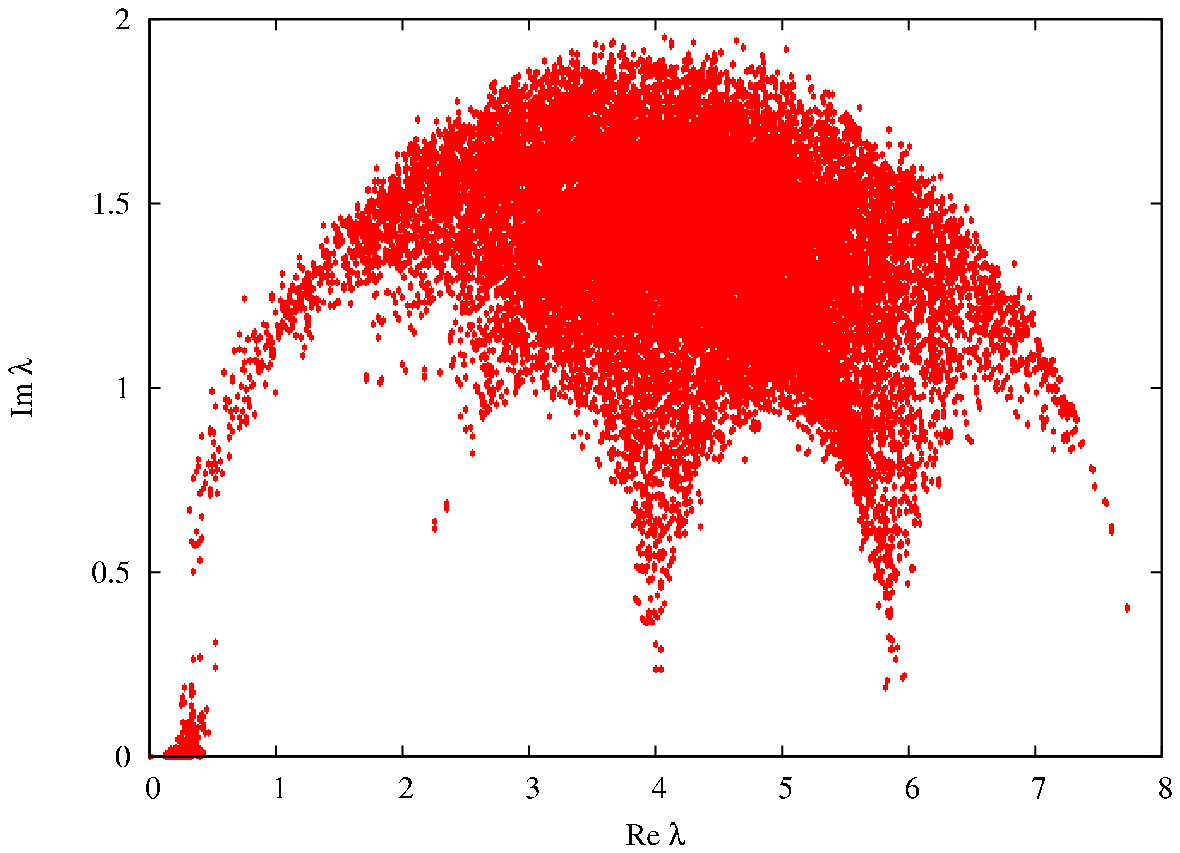}}
\hspace{0.3cm}
\subfigure[$\ \kappa=0.24$]
{\label{fig:dw_N16_b1_k0.24}
\includegraphics[width=7.5cm]{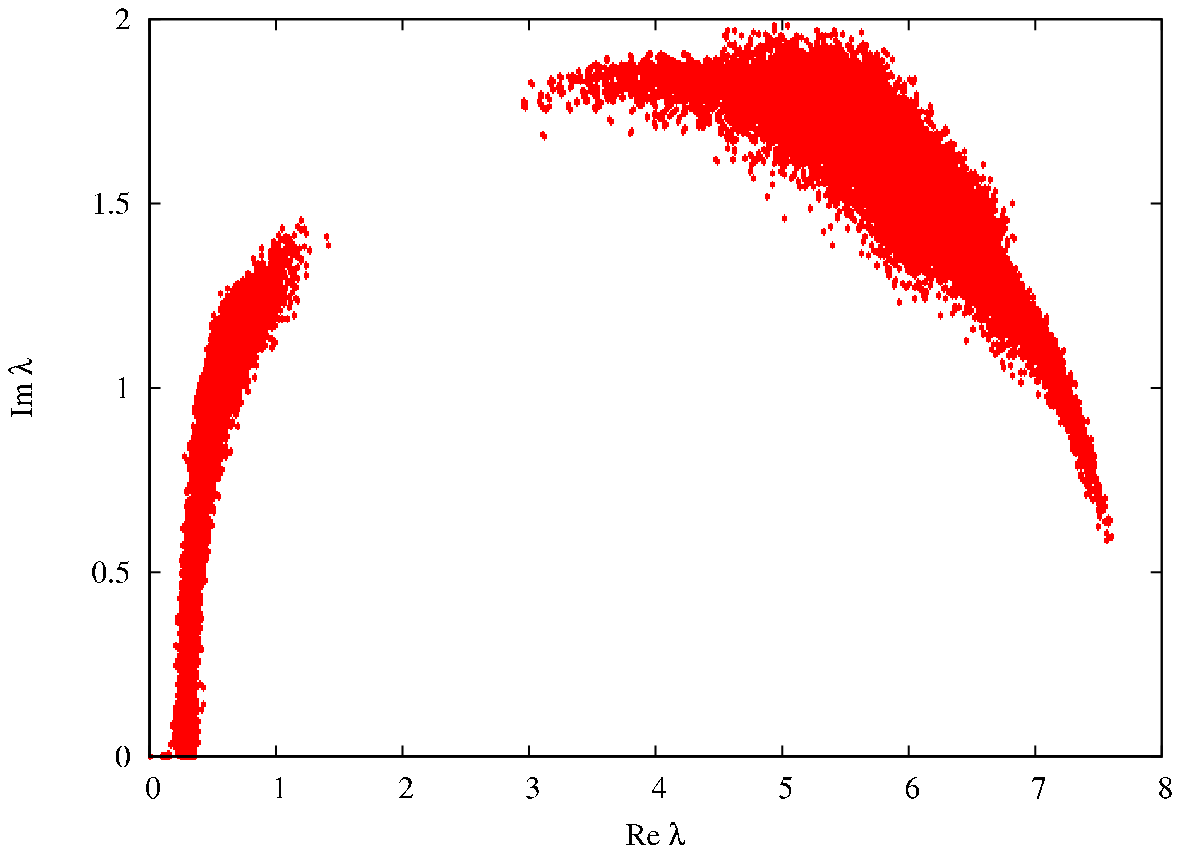}}
\caption{Spectrum of $4D_W(m_0)$ from simulations at $b=1.0$
and $N=30$ at $\kappa=0.01$, $0.03$, $0.12$, $0.14$, $0.17$ and $0.24$.
Only eigenvalues with positive imaginary part are shown.}
\label{fig:DWspectscan}
\end{figure}

In Fig.~\ref{fig:DWspectscan} we show how the spectrum changes
as we vary $\kappa$ at fixed $b=1.0$ and $N=16$. 
At $\kappa=0.01$, where we
are in the $Z_1$ phase (see Figs.~\ref{fig:scan_b1.0}
and \ref{fig:scanM_b1.0}),
we see one main finger and a small indication of a second.
This is consistent with the eigenvalues forming a single clump,
so that the momenta, given by eigenvalue differences, are all small.
At $\kappa=0.03$ we have moved into the $Z_2$ phase, with two clumps
of eigenvalues. We see that this allows the spectrum to spread into
all five fingers, because eigenvalue differences can now range up to
$\pi$. The details of the spectrum differ from those of a large-volume
free fermion, however, in particular having a low density of points in
the central three fingers and a ``rectangular'' shaped envelope.
Nevertheless, it is clear that one must interpret the spectrum
with care---the presence of five fingers alone does not
imply that reduction holds.

The next value, $\kappa=0.12$, is well inside the funnel,
and we see a distribution which is qualitatively similar to that
of a free fermion, with a rounded top and five fingers.
These features are present for all $\kappa<\kappa_c$ inside the
funnel. Particularly noteworthy is
the presence of the comet-shaped clump of eigenvalues near the
origin. We find that there are exactly $4(N-1)$ eigenvalues per configuration
in this clump. We thus interpret them as would-be zero modes, 
i.e. eigenvalues that would be zero if $b\to\infty$.
These modes are dropped in weak coupling calculations, both
because they do not impact the dynamics 
(as they do not depend on the $\theta_\mu^a$)
and because they form only an $O(1/N)$ fraction of the total number
of modes.
The spectrum indicates, however, that they could
have an important impact on the long-distance dynamics which might
overcome their relative paucity.
We recall that for a large-volume Wilson operator
it is the small eigenvalues which determine long-distance
behavior such as chiral symmetry breaking.
For very large $N$ we expect small eigenvalues to come dominantly
from the first finger, which should approach the real axis.
What we see from the figure is that $N=16$ is quite
far from this limit.
Thus we conclude that the would-be zero modes are
a potential source of the $1/N$ corrections observed above
in the plaquette and other quantities,
and that their contribution could be sizable
(given how far the ``true'' low-energy modes in the first finger are
from the real axis).

The spectrum within the funnel but just above the transition is
illustrated by the result for $\kappa=0.14$.
Eigenvalues near $\Re \lambda=0.43$ are suppressed by the determinant.
The would-be zero modes cluster to the left of this excluded point,
while the first finger now approaches closer to the real axis.
The latter feature indicates that the funnel-region above the
transition is more continuum-like, which is consistent with its
larger average plaquette. On the other hand, the spectrum as a whole
is less symmetric about the ${\rm Re}\lambda=4$ line than that below the
transition.

Moving to $\kappa=0.17$, which is still inside the funnel, the would-be
zero modes have spread out again (perhaps because the
excluded point has now moved to $\lambda=1.06$), while the
first finger has become longer and denser. The second finger,
however, has almost disappeared.

Finally, at $\kappa=0.24$ we are in the $Z_3$ phase. 
This is reflected by the spectrum breaking into three distinct regions
(only two being visible since the third has negative imaginary
part), resulting from eigenvalue differences distributed around
$0$ and $\pm 2\pi/3$.

We have done similar scans at lower $b$,
but the results are less illuminating,
because the bulk of the spectrum moves closer to the real axis,
such that, at $b=0.35$, one cannot see any fingers.
A better approach is to use the spectrum of $Q^2$,
as described in the following subsection.

\begin{figure}[tbp!]
\subfigure[$\ N=37$, 300 configs] {\label{fig:dw_N37_b1_k0.12}
\includegraphics[width=7.5cm]{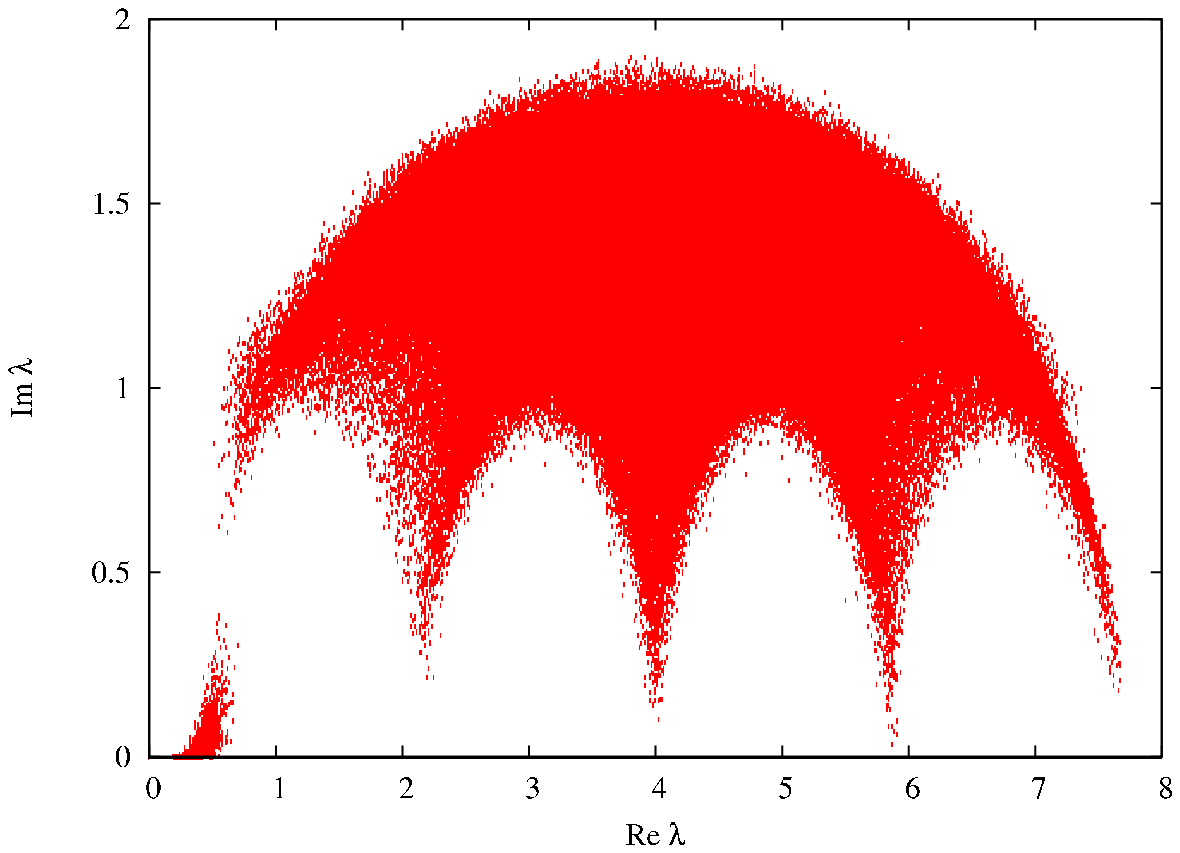}}
\hspace{0.3cm}
\subfigure[$\ N=53$, 150 configs] {\label{fig:dw_N53_b1_k0.12}
\includegraphics[width=7.5cm]{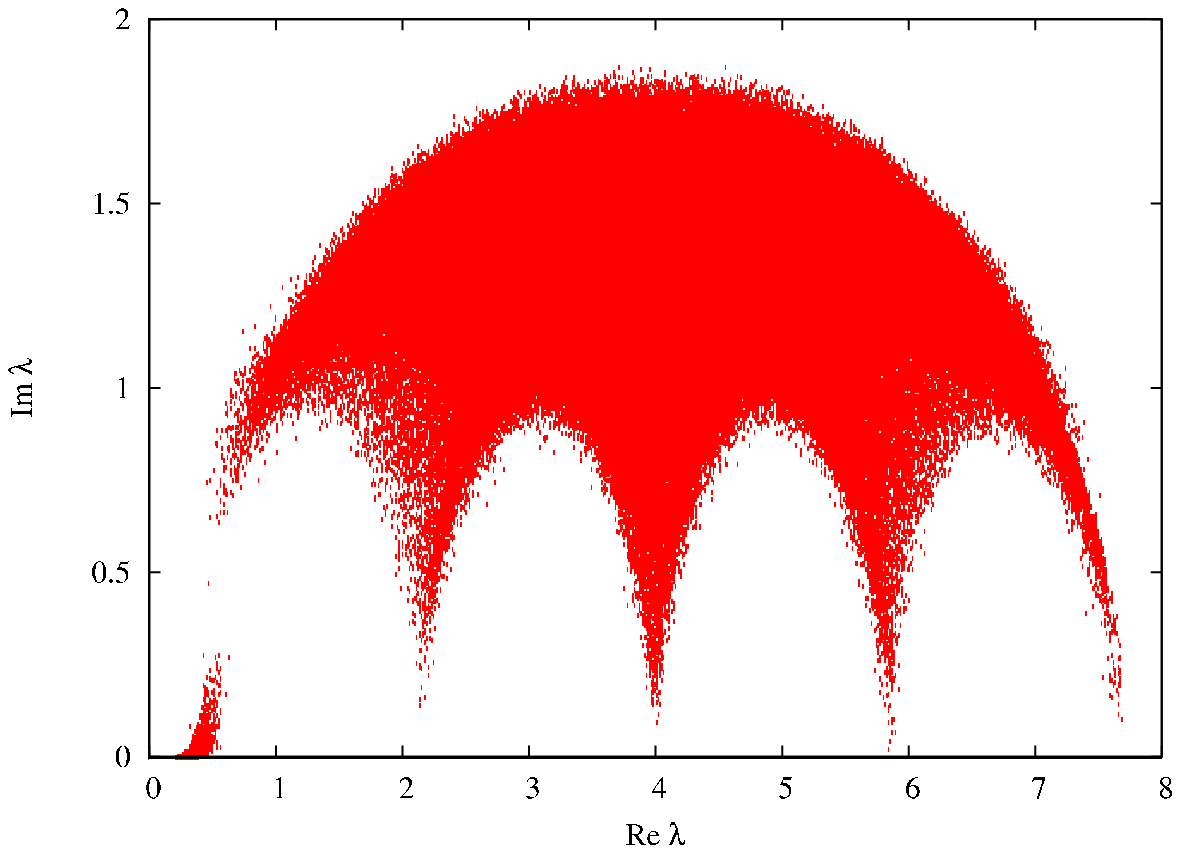}}
\caption{Spectrum of $4D_W(m_0)$ at $b=1.0$ and $\kappa=0.12$, for $N=37$ and
$53$. Note that since $(53/37)^2\approx2$ the number of points is approximately
the same in both plots.}
\label{fig:DWspect_b1_k12}
\end{figure}

We have also studied the $N$ dependence of the spectrum at
$b=1$ and $\kappa=0.12$. Results from $N=37$ and $53$ are shown
in Fig.~\ref{fig:DWspect_b1_k12}, and can be compared to the
$N=16$ results in Fig.~\ref{fig:dw_N16_b1_k0.12}.
The spectra at $N=37$ and $53$ differ very little.
The main changes are that 
the size of the clump of would-be zero-modes decrease as $N$
increases, and that the tip of the first and fifth fingers move down
slightly. The tips of the other fingers, however, barely move.
Compared to $N=16$, on the other hand, the fingers are somewhat
more extended.

We draw several conclusions from these results.
First, the qualitative
agreement of the spectrum within the funnel
with that from a large volume four-dimensional theory 
supports our conclusion that reduction holds therein.
Second, the $\Leff\approx N^{1/4}$,
crystalline distribution of eigenvalues described in the
Appendix is disfavored,
since, for our values of $N$, it would not lead to 
the presence of fingers, and thus differ from our results.\footnote{%
We have studied this further by calculating the spectrum on a
$2^4$ lattice with $N=3$, corresponding to a ``partial crystallization''
of a single-site $N=48$ theory. This spectrum also has no fingers.}
Third, our results are also inconsistent with the
$\Leff\approx N$ model, since the fingertips do not approach
the real axis as fast as the expected $1/N$.
Fourth, the would-be zero-modes are a possible source of $O(1/N)$ corrections.
And, finally, these zero-modes may make an important contribution to
dynamics, thus enhancing the $1/N$ corrections (again,
provided that these $1/N$ corrections are not exactly 
canceled by $1/N$ corrections from the $4(N^2-N)$
modes in the bulk of the eigenvalue distribution).

We have also calculated the spectrum of the Dirac operator in the
fundamental representation. This gives information directly
about the link eigenvalues, rather than their differences.
The results confirm our understanding of the phase diagram and
eigenvalue distributions explained above, but are not shown for
the sake of brevity.

\subsection{Spectrum of $Q^2$}

An alternative view is provided by the spectrum of the 
squared hermitian Wilson-Dirac operator, 
$Q^2=D_W(m_0) D_W(m_0)^\dagger$.
Its eigenvalues, $\lambda_{Q^2}$, are real and positive.
In the continuum limit, the spectrum has a gap, turning on at
$\lambda_{Q^2}=(am_{\rm phys})^2$. Away from the continuum limit, the turn-on
is smoothed, but still begins approximately at the square of physical
bare quark mass~\cite{GiustiLuscher08}.
For small enough quark masses,
and if there is spontaneous chiral symmetry breaking,
the spectrum above the gap is approximately constant,
with a value proportional to the condensate.
Thus the spectrum can teach us about the size of the
quark mass and about long-distance physics (assuming reduction holds).
The information is also contained in the spectrum of $D_W(0)$,
but for $b < 1$, when the fingers are obscured, is hard to extract.
Thus we have used the spectrum of $Q^2$ mainly
for $b=0.35$, which, we recall,
is a bare coupling close to those used in typical large-volume simulations.

\begin{figure}[tbp!]
\includegraphics[width=12cm]{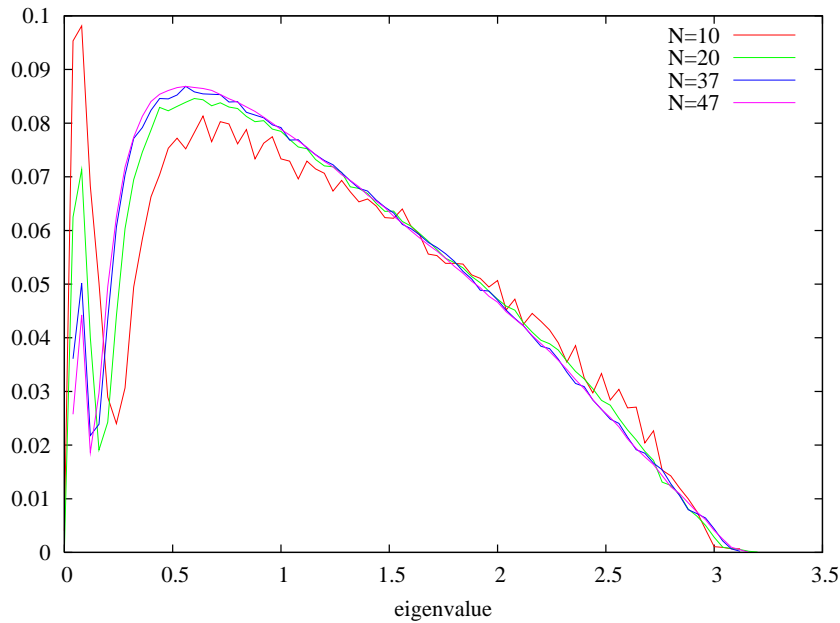}
\caption{Spectrum of $Q^2$ at $b=0.35$ and $\kappa=0.12$,
for $N=10$, $20$, $37$ and $47$, using 150, 150, 20 and 60
configurations, respectively. The vertical scale is arbitrary,
but the relative scales for different $N$ are chosen so that the
area under each spectrum is the same. Errors are not shown,
but can be estimated from the kinks in the spectra.}
\label{fig:Q2spect_b35_k12}
\end{figure}

Results for $b=0.35$, $\kappa=0.12$ and $N=10-47$ are shown in
Fig.~\ref{fig:Q2spect_b35_k12}. (Results at $N=53$ are
very similar to those at $N=47$ but have lower statistics
and are thus not shown.) The spectra are normalized to
have the same integral, so that the large $N$ limit can be taken.
The peak at small eigenvalues has the correct area to contain
just the would-be zero modes, so that its area drops as $1/N$
in the normalized spectrum. It will disappear entirely when
$N\to\infty$. The bulk of the eigenvalues form a ``hump''
which at the upper end ($\lambda_{Q^2}\gtrsim 1.5$) is approximately
independent of $N$, while at the lower end depends on $N$.
The form of the hump at $N=47$ should give a good approximation to
the form at $N=\infty$ because the small-eigenvalue peak has 
little area left to ``redistribute'' to the hump.
A crude extrapolation of the leading edge of the hump at $N=47$
suggests that the gap at $N=\infty$ will be at
$\lambda_{Q^2}\approx 0.1$. This corresponds to a quark mass
of $a m_{\rm phys}\approx 0.3$, modulo the
unknown renormalization factor, which, however, we expect to
be of $O(1)$.
This is the only ``measurement'' of the quark mass that we have
obtained, and shows that the quark is relatively heavy,
not much below $m_{\rm phys} = 1/a$.

These results shed light on the issue of how large $N$ needs to be
for reduction to be useful. On the one hand,
one can seen that the bulk of the eigenvalues (those in the hump)
are close to their large $N$ limit by $N\approx 40$. 
On the other hand, the
would-be zero modes, though making up only an $O(1/N)$ 
fraction of the total,
are the dominant contribution in the low mode region.
Which effect is more important is not clear {\em a priori}---one must
calculate physical observables and study their $N$-dependence.

\subsection{Large Wilson loops}

Our ultimate aim in studying the single-site models is
to use them to calculate physical quantities in phases where
large-N reduction holds.
An important quantity that should be accessible in such phases
is the heavy-quark potential. To obtain this we calculate
rectangular Wilson loops using the large-N reduction recipe
(and also averaging over orientations):
\begin{equation}
W(L_1,L_2) = \frac1{12}\sum_{\mu\ne\nu} \left\langle
 \frac1N {\rm Re}\,\tr\left(
U_\mu^{L_1}U_\nu^{L_2}U_\mu^{\dagger\,L_1}U_\nu^{\dagger\,L_2}\right)
\right\rangle
\,,
\label{eq:strongcouplingW}
\end{equation}
For $N\to\infty$ the result should equal the infinite-volume large-$N$ value.
The potential can be obtained as usual from the large $L_2$ behavior
\begin{equation}
W(L_1,L_2) \stackrel{L_2\to\infty}{\longrightarrow} c(L_1) e^{-V(L_1) L_2}
\,.
\end{equation}
At large $L_1$ we expect linear behavior if we are in a confining regime:
\begin{equation}
\frac{d V(L_1)}{d L_1} \stackrel{L_1\to\infty}{\longrightarrow} \sigma 
\,.
\end{equation}

For finite $N$, reduction will only give useful results
for loops whose sizes satisfy $L_j \ll N$, and the
key question is how much smaller than $N$ do the $L_j$ need to be. 
Another important question is whether it
suffices to calculate Wilson loops using unsmeared links, i.e. whether
the statistical errors will overwhelm any signal of
interest. State-of-the-art calculations in large volumes use 
smearing, as well as other noise reduction techniques.

In our earlier study of the $N_f=1$ model~\cite{AEK}, in which $N\lesssim13$,
we did not find a useful ``window'' as a function of the $L_j$. 
For fixed $L_1$, for example, the
dependence on $L_2$ was a rapid drop for a few points followed by a
slow rise.  The drop was not extensive enough to determine the
potential from the coefficient of the exponential.
Here we have results extending up to $N=53$, and thus expect
that the situation will be substantially better.

\begin{figure}[tbp!]
\includegraphics[width=12cm]{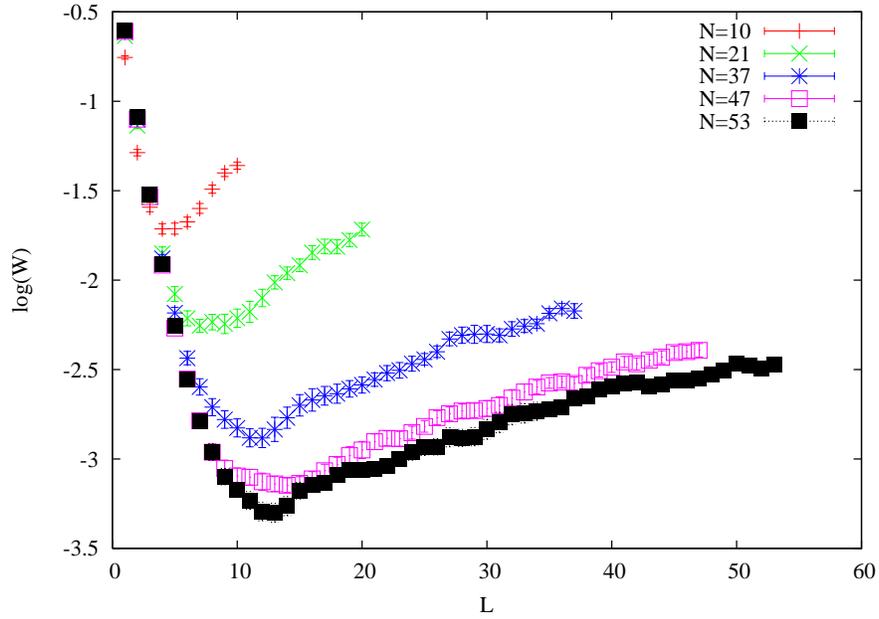}
\caption{Log-linear plot of $1\times L$ Wilson loop versus $L$ for
$L\le N$.
Results are from $b=0.35$, $\kappa=0.12$
and for $N=10$, $21$, $37$, $47$ and $53$,
using $20$ configurations except for $N=10$ where we use 150.}
\label{fig:wlp1xL_b35_k12}
\end{figure}

In Fig.~\ref{fig:wlp1xL_b35_k12} we show results for $1\times L$
loops for a range of values of $N$ on a log-linear scale.
For each value of $N$ we find an approximately exponential decrease
followed by slow, roughly linear increase. The latter we interpret
as a finite $N$ effect, since it begins at larger values of $L$
as $N$ increases. The good news from this plot is that we see convergence
to the expected exponential drop-off as $N$ increases.
For example, at $L=6$ the $N=37$ point has peeled off the linear
envelope, but the $N=47$ and $53$ points are in good agreement.
This is large-$N$ reduction in action.
The bad news is that the maximum
value of $L$ at which convergence occurs, $L_{\rm max}$, 
grows only slowly with $N$.
This is not unexpected: we are trying to extract an exponentially
falling contribution to a quantity which has finite $N$ corrections.
To estimate the size of these corrections, one can look at the
the minimum value of the loop as a function of $L$, which
we find falls as approximately $1/N$.
Thus $L_{\rm max}$ grows only logarithmically
with $N$, which presents a significant numerical challenge.

\begin{figure}[tbp!]
\includegraphics[width=12cm]{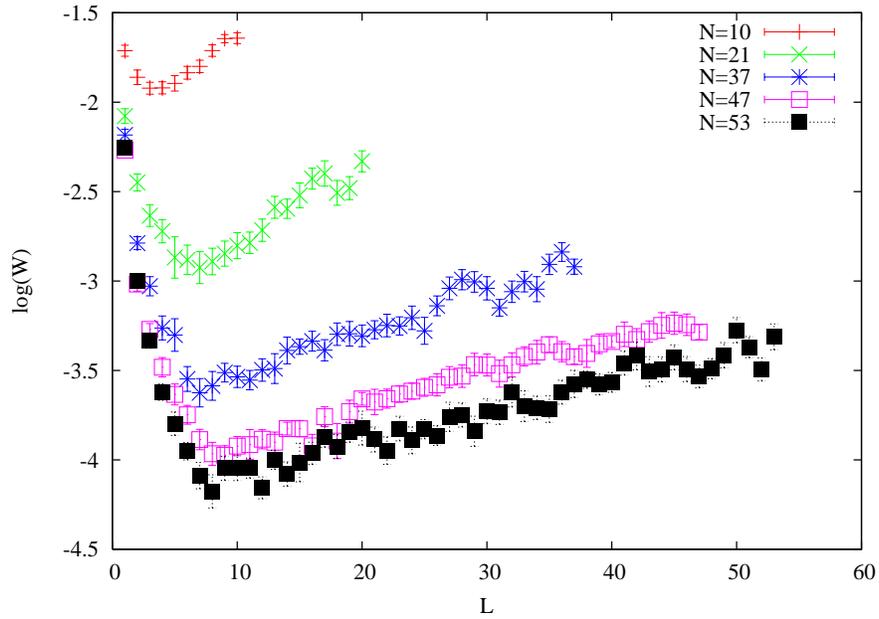}
\caption{As for Fig.~\ref{fig:wlp1xL_b35_k12} except
for $5\times L$ loops.}
\label{fig:wlp5xL_b35_k12}
\end{figure}

Despite this challenge, we see from the figure that we can
extract a value for the slope at small $L$ with reasonably small errors.
This gives $-V(1)$, the potential at unit separation.
To extract $\sigma$, we need the potential at larger separations.
We show in Fig.~\ref{fig:wlp5xL_b35_k12} the results for $5\times L$
loops. The overall pattern is similar to those for 
$1\times L$ loops, but the convergence in $N$ of the falling
parts of the curves is much poorer. Only results for $L\le 2$
appear converged. Thus we cannot extract $V(5)$ with any reliability.
It is of course not a surprise that difficulty of determining
$V(L)$ increases with $L$, since the signal falls off more quickly
while the $1/N$ background is little changed.

Concerning statistical errors, we see from both of these plots that
20 configurations is sufficient to pull out the rapidly falling
part of the curves. The problem is not the statistical errors, but rather
the $1/N$ corrections.

We have carried out a similar investigation at other points in the funnel. 
We find that as $b$ increases, the slope of the initial fall-off decreases.
If we move to the other side of the transition (where the average plaquette
is closer to unity) we find that the slope decreases further, and
also that the large-$L$ approximately linear rise changes to an almost
$L$-independent behavior. 

It would clearly be of interest to gain some understanding
of the large $L$ behavior of the Wilson loops.
The only analytic approach that we are aware of that can shed
some light on the issue is to calculate the loops in strong coupling.
In the $b=0$ limit, and ignoring the fermion determinant, the
gauge links are distributed according to the Haar measure, and
one can show that\footnote{%
Note that these loops all vanish in infinite volume at strong coupling.
One obtains a non-zero result on a single-site lattice
because the $U_\mu$ and $U_\mu^\dagger$ terms in the loop are correlated.}
\begin{equation}
W(L_1,L_2) \stackrel{b,\kappa\to 0}{\longrightarrow}
\frac{1}{N^2-1}\left(L_1 + L_2 - 1 - L_1L_2/N^2\right)
\qquad
\textrm{for}\ 0< L_1,L_2 \le N\,.
\end{equation}
For $L_j\sim O(1)$, the contribution is of $O(1/N^2)$, rising to
of $O(1/N)$ when at least one of the $L_j$ is of $O(N)$. 
This shows how the correlations can lead to an increasing result as
one of the $L_j$ is increased, which is qualitatively in agreement
with the results of Figs.~\ref{fig:wlp1xL_b35_k12} and
\ref{fig:wlp5xL_b35_k12}.
Quantitatively, this model does not, however, reproduce our data at
$b\sim 1$. For example, we would
expect that $W_{\rm min}\sim 1/N^2$ in the model, but we find
instead an approximate $1/N$ dependence.

\section{Discussion and outlook}
\label{sec:outlook}

We have presented a detailed study of the single-site version of 
large-$N$ QCD with two flavors of adjoint Dirac fermions---the so-called 
Adjoint Eguchi-Kawai (AEK) model---discretized using 
Wilson's gauge and fermion actions.
This seemingly simple model turns out to have a rich phase structure, 
as shown in Fig.~\ref{fig:phasediagram}, some aspects of which can
be understood semi-quantitatively~\cite{AHUY}. 
Our most important result is that we find, for $N$ up to $53$,
and for $b$ up to $200$,
a broad funnel in the $(\kappa,b)$ plane in which the $Z_N^4$ center
symmetry is unbroken. This region encompasses both light ($\kappa\to\kappa_c$)
and heavy ($m\sim 1/a$) quarks.
While the funnel narrows as $N\to\infty$, we present
strong evidence that it remains of finite width in this limit.
If so, then the single-site theory, when simulated within the
funnel, is equivalent to the corresponding large-volume theory, 
up to corrections, suppressed by powers of $1/N$, which can be
made arbitrarily small.
Thus one can use the single-site model to study the 
large-$N$ version of the minimal walking technicolor (MWT) model
discussed in the Introduction.

In particular, the phase structure we find {\em within the funnel}
should be identical to that of the large-$N$ MWT model.
We find a first-order transition, with a discontinuity in the
plaquette (and other variables), for values of $b$ at least up
to $b\approx 1$ (see Fig.~\ref{fig:plaq_highb}).
This is in contrast to the results of direct large-volume
simulations of the $N=2$ two-adjoint model, which find
that the transition changes from first-order at strong coupling to
second-order at weak coupling, with the transition occurring at
$b\approx 0.25$~\cite{Catterrall:MWT08,Kari:MWT08}.
The quark mass vanishes along the second-order portion of the
transition line, and it is in part
by studying the non-perturbative $\beta$-function
along this segment of the line 
that evidence has been found for an infrared fixed point.
Further evidence that the $N=2$ theory is conformal in the infrared
comes from studies of the spectrum
and other quantities as one approaches the second-order line
(as summarized in Refs.~\cite{Rummukainen:2011xv,DelDebbio:2011rc}).

The most straightforward interpretation of our phase diagram is
that the theory is confining in the infrared, with
chiral symmetry spontaneously broken, and that the first-order
transition is an example of the first-order scenario of 
Ref.~\cite{SharpeSingleton} in which the chiral condensate flips sign. 
The transition is a result of competition in the effective
potential for the condensate between terms proportional to
the fermion mass $m$ and discretization errors of size $a^2\Lambda^3$.
Here $\Lambda$ is the confinement scale, and it is essential
for this picture that such a scale is present.
The analysis predicts that the discontinuity should
drop rapidly as $b$ increases, 
which is qualitatively consistent with our numerical results.

An alternative interpretation is that the transition is a
``bulk'' transition, which happens to extend to large $b$,
but at some finite $b_c>1$ becomes a second-order line.
This would be an extreme version of what happens for the $N=2$ theory,
where the bulk transition extends to $b\approx 0.25$.
In this picture discretization errors do not allow one to approach
the massless theory, except for extremely weak coupling,
so that one cannot determine the infrared properties 
of the massless theory.\footnote{%
We note that there is no problem in principle with using single-site
models to study theories in the conformal window, as stressed
in Ref.~\cite{UY10}.}
We view this as an unlikely possibility, since bulk transitions
occur generically at strong coupling.


Whatever the interpretation, it is striking that there is such a large
difference in the phase diagrams of the $N=2$ and $N=\infty$ theories. 
On the one hand, such a difference is in conflict
with the models of Ref.~\cite{DietrichSannino}
for the position of the conformal window,
which give N-independent predictions.
On the other, we note
that the beta-function (for the coupling $b$) does depends on $N$,
starting at four-loop order. 
The situation clearly warrants further study.

Another result of our work is that we find strong
evidence for $1/N$ corrections when extrapolating the 
average plaquette to $N=\infty$. 
This is consistent with perturbation theory, which
predicts $1/N$ effects at one-loop order for a single site theory,
in contrast to the $1/N^2$ corrections that one finds in infinite
volume.
Based on our study of the spectrum of $D_W$, the Wilson-Dirac operator, 
we suggest that another source of
$1/N$ corrections may be the contributions of the
$4(N-1)$ modes of $D_W$ that become zero-modes when $b\to\infty$.
One caveat with our suggestion for the source of
the $1/N$ behavior is that the contribution of the would-be zero-modes can
be canceled by those from other modes.
This happens in infinite-volume perturbation theory
(as in the results for the first two terms of the $\beta$-function
mentioned above) and also at strong coupling
[as in the result for Wilson-loops of fixed size
at strong coupling, eq.~(\ref{eq:strongcouplingW})]. 
We also note that the presence of significant $1/N$ dependence in the plaquette
may be related to the apparent difference between the phase
diagrams of the $N=2$ and $N=\infty$ theories.

If, as our results suggest, reduction holds within the funnel,
then a lattice theory with multiple sites is 
being packaged inside the four link matrices and single-site
fermion fields. One can think, approximately, of an
effective lattice size, $\Leff$.
It is important for practical applications to
determine how $\Leff$ grows with $N$.
Our results for the spectrum of $D_W$ (which, within the funnel,
looks qualitatively similar to that on a large volume,
cf. Fig.~\ref{fig:DWspect_b1_k12}) suggest that 
neither the most pessimistic ($\Leff\propto N^{1/4}$) 
nor the most optimistic ($\Leff\propto N$) possibilities hold.
Let us assume, then, that $\Leff\propto N^{1/2}$---an intermediate
possibility motivated in Sec.~\ref{subsec:spectDW}---and
consider the question of whether, if one wants
to determine the large-$N$ properties of a theory, it is
computationally advantageous to use a single-site theory or one on a
large volume, $L^4$. In the former case, we have found
from our simulations that, for fixed $b$ and $\kappa$,
and with $\kappa$ near to $\kappa_c$,
\begin{equation}
{\rm CPU}(\textrm{1-site}) \propto N^{4.5} \propto \Leff^5 N^2
\,,
\end{equation}
where in the final expression we have used $\Leff\propto N^{1/2}$.
For the large volume theory we expect (for fixed lattice spacing
and fermion mass)
\begin{equation}
{\rm CPU}({L^4}) \propto L^5 N^3\,,
\end{equation}
where the $L^5$ is the standard hybrid Monte-Carlo volume
scaling~\cite{HMCvolscaling}, while $N^3$ is the operation
count for the core operation of $N\times N$ matrix multiplies.
This comparison suggests that
the single-site approach could be computationally advantageous.
While there are many caveats to this conclusion
(e.g. the single-site scaling form is based on simulations for
a finite range of $N$, and the scaling
of $\Leff$ with $N$ is not established),
we take it as motivation to further pursue studies of reduced models.

One aspect of such studies is calculating physical quantities
such as the string tension and particle masses. 
We have taken a first step in this direction by calculating
large Wilson loops and attempting to extract the heavy-quark
potential. We find that we can do so for small separations
(roughly out to 3 lattice spacings for $b=0.35$) but not beyond.
The difficulty arises because the signal must be determined from an
exponential decay as a function of the loop size, while
the corrections to reduction lead to a ``background'' of
$O(1/N)$ which is approximately independent of loop size.
As the coefficient of the decay---the potential, or more generally
a hadron mass---increases, one has to go to ever higher values of $N$.
This problem should be less serious, however, for light particles,
such as one expects to find as $\kappa$ approaches $\kappa_c$.
Indeed, calculating the pion mass would allow an important
cross-check on our preferred interpretation that the system
is confining, and chiral-symmetry breaking, in the infrared.

One can also use the $N_f=2$ AEK model away from the
critical line as a single-site model whose long distance
physics is that of the pure gauge theory. In other words,
heavy adjoint fermions resolve the problems of
the original Eguchi-Kawai model. The same holds true for the $N_f=1$ 
theory~\cite{AEK,AHUY}.

To address the unresolved issues described above, one will need
either to work at larger $N$ or move to models with more than one site.
The latter option seems most practical, and also has the advantage of
being simpler to parallelize. First steps in this direction have
been taken in Ref.~\cite{CGU}.
It may also be advantageous to use twisted boundary conditions,
as has been done for the $N_f=1$ theory in Ref.~\cite{AHUY},
since these appear to reduce the power of the corrections from
$1/N$ to $1/N^2$.
One can also consider using improved gauge and fermion actions,
since these are known to clarify the infrared behavior 
in large-volume simulations~\cite{dGSS11}.

\section*{Acknowledgments}

We thank Mithat \"Unsal for discussions and 
Adi Armoni, Simon Catterall, Ari Hietanen, 
Mithat \"Unsal and the referee for comments on the
manuscript. This work was supported in part by the U.S. DOE
Grant No. DE-FG02-96ER40956, and
by Foundation for Polish Science MPD Programme co-financed 
by the European Regional Development Fund, agreement no. MPD/2009/6. 
MK is grateful to the University of Washington for hospitality. 
Most of the numerical simulations were done using the 
Shiva computing cluster at the Faculty of Physics, 
Astronomy and Applied Computer
Science, Jagiellonian University, Cracow.

\appendix

\section{Models for eigenvalues of $D_W$}
\label{sec:app}

In this appendix we describe various possible behaviors of the
link eigenvalues and their implications for the eigenvalues of $D_W$.
These models guide the interpretation of the results presented
in Sec.~\ref{subsec:spectDW} for the spectrum of $D_W$.

We consider the extreme weak coupling
limit, $b\gg 1$, in which we must choose links that maximize $S_{\rm gauge}$. 
This is achieved by links which can simultaneously diagonalized 
by a gauge transformation, i.e. for which one can have
\begin{equation}
U_\mu = {\rm diag}\left(
e^{i\theta_\mu^1},e^{i\theta_\mu^2},\dots e^{i\theta_\mu^N}\right)
\ \ \forall \mu\,.
\end{equation}
What is needed for $D_W$ is the link in the adjoint representation
[see Eq.~(\ref{eq:D_W})].
It is convenient to add a singlet and consider
the link in the reducible $N\otimes\overline{N}$ representation.
In this case, it has composite indices, $A=(a_1,a_2)$, with $a_j=1,N$,
and is also diagonal:
\begin{equation}
\left(U_\mu^{N\otimes\overline{N}}\right)_{AB}
=\left(U_\mu\right)_{a_1 b_1} \left(U_\mu\right)_{a_2 b_2}
\ \ \Rightarrow\ \
U_\mu^{N\otimes\overline{N}} = {\rm diag}
\left(\dots,e^{i(\theta_\mu^{a_1}-\theta_\mu^{a_2})},\dots\right)
\,.
\end{equation}
Inserting this into the massless Wilson-Dirac operator,  
one finds\footnote{%
We multiply by $4=1/[2\kappa_c(g^2=0)]$ in order that
(in large volume) the operator becomes the Dirac operator
with standard normalization in the naive continuum limit.
This undoes the standard renormalization of the fermion fields
by $\sqrt{2\kappa}$ that is used to write $D_W$ in the
form of Eq.~(\ref{eq:D_W}).}
\begin{equation}
4 D_W(m_0=0) = {\rm diag}
\left(\dots,
\left\{(4-\sum_\mu\cos\theta_\mu^{a_1 a_2})
       + i \sum_\mu \gamma_\mu \sin\theta_\mu^{a_1 a_2}
       \right\},\dots\right)
\,, \label{eq:DWm_0}
\end{equation}
where we are using the abbreviation
\begin{equation}
\theta_\mu^{a_1 a_2} = \theta_\mu^{a_1}-\theta_\mu^{a_2}
\,.
\end{equation}
Thus $D_W$ is diagonal in color space, but not in Dirac space.
The eigenvalues of $D_W(m_0=0)$ are as follows. There are $4 N$ 
zero-modes, occurring when $a_1=a_2$ so that
$\theta_\mu^{a_1 a_2}=0$. Four of these are from the singlet, 
which we can now remove, leaving $4(N-1)$ from the adjoint.
The remaining $4N(N-1)$ are each doubly degenerate
(due to charge conjugation symmetry), come in complex conjugate pairs
(due to $\gamma_5$-hermiticity), and have values
\begin{equation}
\lambda_{a_1 a_2} = (4-\sum_\mu\cos\theta_\mu^{a_1 a_2})
\pm i \sqrt{\sum_\mu\sin^2\theta_\mu^{a_1 a_2}}\,,\quad
(a_1\ne a_2)\,.
\label{eq:DWevalues}
\end{equation}

The form of $D_W$ in Eq.~(\ref{eq:DWm_0})
is exactly that of a four-dimensional free massless Wilson-Dirac fermion,
with the momenta in lattice units $a p_\mu$ replaced by $\theta_\mu^{a_1 a_2}$.
This is the standard way in which
the large volume appears in the weak coupling limit,
i.e. link eigenvalue differences become momenta~\cite{BHN}. 
If these eigenvalue
differences are distributed such that the resulting ``momenta''
lie uniformly throughout a four-dimensional Brillouin zone, then
the single-site $D_W$ will approximate that of large-volume four-dimensional
theory. The resulting spectrum has the five fingers mentioned above,
whose ``tips'' occur when all four $a p_\mu$ equal $0$ or $\pi$.
The tips are distinguished
by the number of $a p_\mu$ which equal $0$---either $0$, $1$, $2$,
$3$ or $4$.
Alternatively, if the eigenvalues are correlated in some way,
then the spectrum will not, in general, have five fingers.
There would, for example, be fewer fingers if the effective dimensionality
is less than four.\footnote{%
Given the weak-coupling form (\ref{eq:DWevalues}), the spectrum
is necessarily confined to lie between the ellipse $(R-4)^2 + (2I)^2 = 16$
and the four circles $(R-R_0)^2+I^2=1$, with $R_0=1$, $3$, $5$, and $7$.
Here $R$ and $I$ are respectively the real and imaginary parts of the
eigenvalues. Thus the distribution is kinematically
forced to lie in one of the five fingers once $|I|<1$. The presence
of fingers {\em per se} is thus not significant, but the
number of fingers which are populated is significant.}

\begin{figure}[tbp!]
\includegraphics[width=12cm]{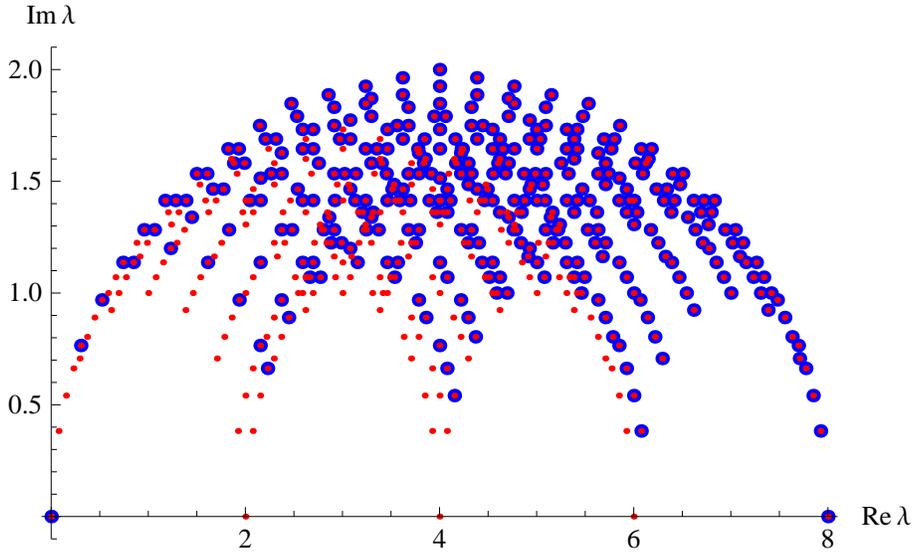}
\caption{Spectrum of $4D_W(0)$ in the model given by
Eq.~(\ref{eq:model1}) for $N=16$ and with $O(N^2)$ configurations
(large [blue] dots) compared to the spectrum of the
free Wilson-Dirac operator on a $16^4$ lattice (small [red] dots).
Only eigenvalues with positive imaginary part are shown;
the spectra are symmetric under reflection in the ${\rm Re}\lambda$
axis.}
\label{fig:freespect_N16}
\end{figure}

To illustrate these comments we discuss the results from three
simple (and somewhat artificial) models for the eigenvalue distributions.
In the first, we choose the eigenvalues in each direction
to be evenly spaced around the unit circle,
but in a randomly permuted order, with the permutation being
independent in each direction. 
This leads to
\begin{equation}
\theta_\mu^{a_1 a_2} = \frac{2\pi}{N} [\sigma_\mu(a_1)-\sigma_\mu(a_2)]
\,,
\label{eq:model1}
\end{equation}
with $\sigma_\mu$ a permutation of $1-N$.
In the eigenvalue-momentum correspondence these are the 
subset of the momenta available on an $N^4$ lattice
with periodic boundary conditions. The Brillouin zone of
such a lattice contains $N^4$ momenta, which is much larger than
the $N^2$ values of $\theta_\mu^{a_1 a_2}$ produced by a
single configuration.
We thus assume further that independent configurations lead
to independent permutations. Then, with of $O(N^2)$ configurations,
one obtains the spectrum shown in Fig.~\ref{fig:freespect_N16}
by the large (blue) dots. This is compared in the figure to
the full spectrum of a 4-d free Wilson-Dirac operator on an $N^4$ lattice.
One sees the appearance of the desired 5 fingers, but also that
some points in the full spectrum are missing. This is due to the
fact that $\theta_\mu^{a_1 a_2}=0$ only if
$a_1=a_2$, in which case $\theta^{a_1 a_2}_\mu=0$ for all $\mu$.
The model thus cannot produce momenta proportional to $(0,0,0,n_4)$,
$(0,0,n_3,n_4)$ or $(0,n_2,n_3,n_4)$ (or their permutations).
These ``missing modes'' have the largest impact on the left-most
(and thus physical) finger, but become increasingly unimportant
as $N$ increases.
We note that the distance of the fingertips to the real axis
scales as $1/N$.

In this model center symmetry is unbroken, and in particular
all traces of open loops, $K_{n}$, vanish 
(unless all four $n_\mu$ are integer multiples of N).
It gives an example where the effective size is $\Leff=N$,
in the sense that the fermion operator after averaging over
configurations has the same spectrum as a theory with volume $N^4$.
The model is artificial in that eigenvalues do not fall on
the ``clock'' values in unbroken phase, but,
as seen above, are spread nearly uniformly.
Nevertheless, it indicates how the spectrum of $D_W$ can teach
us about the distribution of eigenvalues and the effective
dimensionality.

Our second model is a variant of the first in which the eigenvalues
still take clock values, but they are fully correlated between the
directions. This breaks the $Z_N^4$ center symmetry down to the
``diagonal'' $Z_N$ subgroup.
More precisely, we assume that
$\theta_\mu^a = \theta_\nu^a + (2\pi/N) n_{\nu,\mu}$ for all $a$,
$\mu$ and $\nu$, with $n_{\nu,\mu}$ an integer.
This leads to 
\begin{equation}
\theta_\mu^{a_1 a_2} = \frac{2\pi}{N} [\sigma(a_1)-\sigma(a_2)]
\,,
\label{eq:model2}
\end{equation}
where now there is a single permutation $\sigma$ for all four directions.
This is the type of ``locking'' found in the quenched EK (QEK) 
model~\cite{QEKBS}.
It leads to the spectrum of $D_W$ being that of a 1-d free Wilson
fermion in a periodic box of length $\Leff=N$, scaled up by
a factor of 4.
In this case the central three fingers are missing.

The third model is inspired by the analysis of
Ref.~\cite{UY10}, in which it is shown that, at extremely
weak coupling, the repulsion
between eigenvalues leads to the formation of a four-dimensional
crystal if $K=N^{1/4}$ is an integer, and an approximately
uniform distribution for other values of $N$.
It is important to note that the analysis of Ref.~\cite{UY10} holds
only if the coupling $b$ evaluated at the scale $1/(a \Leff)$ is much
larger than unity. This requires that the lattice coupling,
$b(1/a)$, grows logarithmically with $N$. This is not the
standard limit in which reduction holds, in which $b$ is fixed.
Indeed, volume independence does not hold in this regime.
We can express this realization of eigenvalues as 
\begin{equation}
\left[\theta_1^a,\theta_2^a,\theta_3^a,\theta_4^a\right]
=
\frac{2\pi}{K}
\left[{\rm mod}(a\!-\!1,K),{\rm mod}(\frac{a\!-\!1}K,K),
       {\rm mod}(\frac{a\!-\!1}{K^2},K),
       {\rm mod}(\frac{a\!-\!1}{K^3},K)\right]
\,,
\label{eq:model3}
\end{equation}
which breaks the center symmetry down to $Z_K^4$, but it
is argued in Ref.~\cite{UY10} that fluctuations in the
eigenvalues can lead to an averaging over the different
``crystals'' related by $Z_N^4$ transformations, and
thus the restoration of the full symmetry.
The momenta $\theta^{a_1,a_2}_\mu$ are the same for all
crystals, and lead to a four-dimensional spectrum for $D_W$.\footnote{%
Note that in this case a single configuration suffices
to fill out the spectrum.}
The difference from the first model is that instead of
the spectrum being that on a lattice of size $\Leff=N$ 
it is on the much smaller size $\Leff=K=N^{1/4}$.
This is the same scaling as for the orbifold construction
discussed in the main text, and is the most conservative possibility.

For the values of $N$ that we use, one has $\Leff<3$.
For such a small lattice, even though it is four-dimensional,
the spectrum shows no fingers, i.e. no eigenvalues close to
the real axis. 
Thus if this model of the eigenvalues
provides even an approximate description of our data, we would
not expect to see fingers.

It is interesting that the above distribution of eigenvalues was first 
suggested in the context of the space embedding into color space of the 
QEK model~\cite{GK}, and then analyzed in \Ref{Bars}. 
For a further discussion on this point see \Ref{QEKBS} where two 
of us analyzed this eigenvalue distribution within the QEK model 
(where it is referred to as the `Brillouin Zone' distribution).

\end{document}